\documentclass [a4paper]{article}

\usepackage[hiresbb]{graphicx}

\usepackage{setspace}
\usepackage{float}
\usepackage[title]{appendix}
\usepackage{nameref,hyperref}
\usepackage[table]{xcolor}
\usepackage{seqsplit}
\usepackage{lscape}
\usepackage{longtable}
\usepackage{rotating}
\usepackage{caption}
\usepackage{subcaption}
\usepackage{amsmath}
\usepackage{bm}
\usepackage{lineno}

\begin{document}
\fontsize{10.5pt}{18pt}\selectfont

\fontsize{13pt}{18pt}\selectfont
\begin{center}
{\bf Do economic effects of the anti-COVID-19 lockdowns in different regions\\ interact through supply chains?
}

\vspace{2ex}

\fontsize{11pt}{18pt}\selectfont
Hiroyasu Inoue\footnote{Graduate School of Simulation Studies, University of Hyogo, inoue@sim.u-hyogo.ac.jp.}
Yohsuke Murase\footnote{RIKEN Center for Computational Science, yohsuke.murase@gmail.com}
Yasuyuki Todo\footnote{Graduate School of Economics, Waseda University, yastodo@waseda.jp.}
\vspace{3ex}

\fontsize{10.5pt}{18pt}\selectfont
Abstract
\end{center}

\vspace{1ex}
\fontsize{10pt}{12pt}\selectfont

\noindent To prevent the spread of COVID-19, many cities, states, and countries have `locked down', restricting economic activities in non-essential sectors. Such lockdowns  have substantially shrunk production in most countries. This study examines how the economic effects of lockdowns in different regions interact through supply chains, a network of firms for production, simulating an agent-based model of production on supply-chain data for 1.6 million firms in Japan. We further investigate how the complex network structure affects the interactions of lockdowns, emphasising the role of upstreamness and loops by decomposing supply-chain flows into potential and circular flow components. We find that a region's upstreamness, intensity of loops, and supplier substitutability in supply chains with other regions largely determine the economic effect of the lockdown in the region. In particular, when a region lifts its lockdown, its economic recovery substantially varies depending on whether it lifts lockdown alone or together with another region closely linked through supply chains. These results propose the need for inter-region policy coordination to reduce the economic loss from lockdowns.

\vspace{1ex}





\newpage

\fontsize{10pt}{10pt}\selectfont

\section{Introduction}

COVID-19, a novel coronavirus (SARS-CoV-2) disease, has been spreading worldwide. To prevent its spread, many cities, regions, and countries were or have been `locked down', suppressing economic activities. On 18 April 2020, 158 countries out of 181 required closing temporarily or working from home for some sectors in some or all cities. Although some countries later lifted their lockdowns, the number of countries in a lockdown remained 95 on 30 July 2020~\cite{Hale2020}.

Closing workplaces shrinks the economic output of locked down regions. The negative economic effect of a lockdown in one region may diffuse through supply chains, i.e., supplier-client relationships of firms, to other regions that are not necessarily locked down. When a firm is closed by a lockdown strategy, its client firms located anywhere should suffer decreased production because of the lack of supply of intermediate goods and services. Suppliers of the closed firm should also see reduced production because of a shortage of demand. 

Many studies have empirically confirmed the propagation of economic shocks through supply chains, particularly shocks originating from natural disasters~\cite{Barrot2016, Boehm2019, Carvalho2016, Inoue2019, Inoue2019b, Kashiwagi2018}. Some examine the diffusion of the effect of lockdowns because of COVID-19 on production across regions and countries and estimate the total effect using input--output (IO) linkages at the country-sector level~\cite{Bonadio2020, Guan2020, McCann2020, McKibbin2020} and supply chains at the firm level~\cite{Inoue2020}.

Several works focusing on natural disasters~\cite{Inoue2019, Inoue2019b} pay attention to how the network structure of supply chains affects the propagation of shocks, finding that the scale-free property, non-substitutability of suppliers, and loops are major drivers of such propagation. However, the role of the network structure has not been fully examined in the context of the propagation of the lockdown effect. However, this issue should be of great interest from the perspective of network science for the following two reasons. 

First, the literature on network interventions has investigated what types of individuals or groups in a network, such as those with high centrality, should be targeted to promote (prevent) the diffusion of positive (negative) behaviours and outcomes~\cite{Valente2012, Valente2017}. Similarly, we are interested in how the economic effect of imposing and lifting a lockdown in one region, an example of a network intervention, diffuses to other regions. Compared with existing works, this study is novel in many respects. For example, we consider interventions in a network of firms and their economic outcomes, while previous studies focus on the health behaviours and outcomes in human networks~\cite{Hunter2019}, with a few exceptions that examine economic outcomes in human networks~\cite{Sciabolazza2020}. In addition, because a lockdown is usually imposed in a city, state, or country, the scale of interventions is large. Those firms targeted by such interventions are exogenously determined by geography, and thus we should assess the network characteristics of exogenously grouped nodes rather than the endogenously connected ones identified by network centrality~\cite{Valente2012, Robins2005} or community detection algorithms~\cite{Newman2004}. 

Second, at any point amid the spread of COVID-19, some regions have imposed a lockdown, while others have remained open. Therefore, when we evaluate the lockdown strategy of a region, the interactions between the strategies of different regions need to be taken into account. In other words, the economic effect of a lockdown in a region depends on whether other regions connected through supply chains are similarly locked down. For example, Sweden did not impose a strict lockdown, unlike other European countries. However, it still expects a 4.5\% reduction in gross domestic product (GDP) in 2020, a decline comparable to those in neighbouring countries that did lock down, possibly because of its close economic ties with its neighbours~\cite{NYT2020}. Motivated by the Swedish experience, this study examines the network structure between regions that is usually ignored in the literature on network interventions and discusses the need for policy coordination among regions depending on their network characteristics. Some studies call for inter-regional and international policy coordination in the presence of spillover effects in the context of health, environment, and macroeconomics~\cite{Kremer2007, Taylor2013}, but they do not explicitly incorporate the network structure. 

We conduct a simulation analysis applying actual supply-chain data of 1.6 million firms and experiences of lockdowns in Japan to an agent-based model of production. Specifically, we analyse the network characteristics of a prefecture in Japan that led to greater economic recovery by lifting its lockdown when all other prefectures remained locked down. In addition, to further highlight the interactions between regions, our simulation investigates how the characteristics of the supply-chain links between two prefectures affect their economic recovery when they simultaneously lift their lockdowns. One novelty of our study is to decompose supply-chain flows into potential and loop flow components and test the role of upstreamness (potential) in supply chains and intra- and inter-prefectural loops in diffusion.


\section{Data}

The data used in this study are taken from the Company Information Database and Company Linkage Database collected by Tokyo Shoko Research (TSR), one of the largest credit research companies in Japan. The former includes information about the attributes of each firm, including the location, industry, sales, and number of employees, whereas the latter includes the major customers and suppliers of each firm. Because of data availability, we use the data on firm attributes and supply chains in 2016. The number of firms in the data is 1,668,567 and the number of supply-chain links is 5,943,073. Hence, our data identify the major supply chains of most firms in Japan, although they lack information about supply-chain links with foreign entities. Because the transaction value of each supply-chain tie is not available in the data, we estimate sales from a supplier to each of its customers and consumers using the total sales of the supplier and its customers and 2015 Input-Output Tables for Japan~\cite{METI15}. In this estimation process, we must drop firms without any sales information. Accordingly, the number of firms in our final analysis is 966,627 and the number of links is 3,544,343. Although firms in the TSR data are classified into 1,460 industries according to the Japan Standard Industrial Classification~\cite{MIC2013}, we simplify them into the 187 industries classified in the IO tables. Supplementary Information~\ref{ch:appdata} provides the details of the data construction process.

In the supply-chain data described above, the degree, or the number of links, of firms follows a power-law distribution~\cite{Inoue2019}, as often found in the literature~\cite{Barabasi16}. The average path length between firms, or the number of steps between them through supply chains, is 4.8, indicating a small-world network. Using the same dataset, previous studies~\cite{Inoue2019, Fujiwara10} find that 46--48\% of firms are included in the giant strongly connected component (GSCC), in which all firms are indirectly connected to each other through supply chains. The large size of the GSCC clearly shows that the network has a significant number of cycles unlike the common image of a layered or tree-like supply-chain structure.

\section{Methods}

\subsection{Model}
\label{ch:model}

Agent-based models that incorporate the interactions of agents through networks have been widely used in social science recently~\cite{Axtell2006, LeBaron2008, Gomez2019}. Following the literature, we employ the dynamic agent-based model of Inoue and Todo~\cite{Inoue2019,Inoue2019b}, an extension of the model of Hallegatte~\cite{Hallegatte08} that assumes supply chains at the firm level. In the model, each firm utilises the inputs purchased from other firms to produce an output and sells it to other firms and consumers. Firms in the same industry are assumed to produce the same output. Supply chains are predetermined and do not change over time in the following two respects. First, each firm utilises a firm-specific set of input varieties and does not change the input set over time. Second, each firm is linked with fixed suppliers and customers and cannot be linked with any new firm over time, even after a supply-chain disruption. Accordingly, our analysis should viewed as that focusing on short-term changes in production. Furthermore, we assume that each firm keeps inventories of each input at a level randomly determined from the Poisson distribution. Following Inoue and Todo~\cite{Inoue2019}, in which parameter values are calibrated from the case of the Great East Japan earthquake, we assume that firms aim to keep inventories for 10  days of production on average (see Supplementary Information~\ref{ch:appmodel} for the details).

When a restriction is imposed on the production of a firm, both the upstream and the downstream of the firm is affected. On the one hand, the demand of the firm for the parts and components of its suppliers immediately declines, and thus suppliers have to shrink their production. Because demand for the products of suppliers' suppliers also declines, the negative effect of the restriction propagates upstream. On the other hand, the supply of products from the directly restricted firm to its customer firms declines. Then, one way for customer firms to maintain the current level of production is to use their inventories of inputs. Alternatively, customers can procure inputs from other suppliers in the same industry already connected before the restriction if these suppliers have additional production capacity. If the inventories and inputs from substitute suppliers are insufficient, customers have to shrink their production because of a shortage of inputs. Accordingly, the effect of the restriction propagates downstream through supply chains. Such downstream propagation is likely to be slower than upstream propagation because of the inventory buffer and input substitution.  

\subsection{Lockdowns in Japan}
\label{sec:soe}

In Japan, lockdown strategies were implemented at the prefecture level under the state of emergency~\cite{Cabinet2020} first declared on 7 April 2020 in seven prefectures with a large number of confirmed COVID-19 cases. Because populated regions tended to be affected more and earlier, these seven prefectures were industrial clusters in Japan, including Tokyo, Osaka, Fukuoka and their neighbouring prefectures. The target of the state of emergency was expanded to all 47 prefectures on 16 April. The state of emergency was lifted for 39 prefectures on 14 May and for an additional three on 21 May; it was lifted for the remaining five prefectures on 25 May. Supplementary Information~Figure~\ref{fig:mapsoe} summarises the timeline of the lockdowns in different prefectures. 


Although the national government declared a state of emergency, the extent to which the restrictions were imposed was determined by each prefecture government. Therefore, the level of the lockdown in each prefecture may have varied. Although all prefectures were in the state of emergency from April 16 to May 14, prefectures with a large number of confirmed COVID-19 cases, such as the seven prefectures in which the state of emergency was first declared, requested more stringent restrictions than others. The national or prefecture government can only request closing workplaces, staying at home, and social distancing rather than require these actions through legal enforcement or punishment; however, strong social pressure in Japan led people and businesses to voluntarily restrict their activities to a large extent. As a result, production activities including those in sectors not officially restricted shrunk substantially (Mainichi Newspaper, 27 May 2020).
 
\subsection{Simulation procedure}
\label{sec:methodsim}

\paragraph*{Replication of the actual effect} In our simulation analysis, we first confirm whether our model and data can replicate the actual reduction in production caused by the lockdown of Japan during the state of emergency. Because we cannot observe the extent to which each firm reduces its production capacity by obeying governmental requests, the rate of reduction in production capacity for each sector assumed in our simulation analysis depends on its characteristics. Because the reduction rate particularly during the lockdowns in Japan is not available, we follow the literature that defines the reduction rate in general settings. Specifically, the rate of reduction in a sector is the product of the level of reduction determined by the degree of exposure to the virus given by~\cite{Guan2020} and the share of workers who cannot work at home given by~\cite{Bonadio2020}. For example, in lifeline/essential sectors such as the utilities, health, and transport sectors, the rate of reduction is assumed to be zero; in other words, the production capacity in these sectors does not change on lockdown. In sectors in which it is assumed that exposure to the virus is low (10\%) and 47.5\% of workers can work at home, such as the wholesale and retail sectors, the rate of reduction is 5.25\% ($=0.1 \times (1-0.475)$). Sectors with medium exposure (50\%) and a lower share of workers working at home (26.8\%), such as the iron and other metal product sectors, reduce production capacity by 35.2\% ($=0.5 \times (1-0.268)$). See Supplementary Information~Table~\ref{tbl:mul} for the rate of reduction of each sector.

After the lockdown in a prefecture is lifted, all the firms in that prefecture immediately return to their pre-lockdown production capacity. Moreover, we assume that inventories do not decay over time: inventories stocked before the lockdown can be fully utilised after the lockdown is lifted. The following results are averaged over 30 Monte Carlo runs. 

\paragraph*{Interactions among regions} After checking the accuracy of our simulation model, we examine how changing the restriction level of the lockdown in a region affects production in another region. For this purpose, we experiment with different sets of sector-specific rates of reduction in production capacity by multiplying the benchmark rates of reduction defined above by a multiplier such as 0.4 or 0.8. For example, when the benchmark rate of reduction in a sector is 35.2\%, as in the case of the iron and other metal product sectors, and the multiplier is 0.4, we alternatively assume a rate of reduction of 14.1\%.

Moreover, we assume that the rates of reduction can vary among prefectures because each prefecture can determine its own level of restrictions under the state of emergency (Section \ref{sec:soe}). In practice, the restrictions requested by the prefecture government were tougher and people were more obedient to the requests in the seven prefectures in which the state of emergency was first declared because of their large number of confirmed COVID-19 cases (Figure~\ref{fig:mapsoe}(b)) than in other prefectures. Accordingly, we run the same simulation assuming different rates of reduction for the two types of prefectures, defined as more and less restricted groups, to investigate how different rates of reduction in one group affect production in the other. 

\paragraph*{Lifting lockdown in only one region} In practice, some prefectures lifted their lockdowns earlier than others (Section \ref{sec:soe}). Although this may have led to the recovery of value added production, or gross regional product (GRP), the extent of such a recovery should have been affected by the links between firms in the prefecture and others still locked down. To highlight this network effect, we simulate what would happen to the GRP of a prefecture if it lifted its lockdown while all others were still imposing lockdowns. Then, we investigate what network characteristics of each prefecture determine the recovery from lockdown, measured by the ratio of the increase in the GRP of the prefecture by lifting its lockdown to the reduction in its GRP because of the lockdown of all prefectures. 

In particular, we focus on four types of network characteristics. First, when a prefecture is more isolated from others in the supply-chain network, the effect of others' lockdowns should be smaller. We measure the level of isolation using the number of links within the prefecture relative to the total degree of firms (total number of links from and to firms) in the prefecture.

Second, an alternative and more interesting measure of isolation is the intensity of loops in supply chains. Although supply chains usually flow from suppliers of materials to those of parts and components and to assemblers, some suppliers use final products such as machinery and computers as inputs. This results in many complex loops in supply chains \cite{kichikawa2019community}, in which negative shocks circulate and can become aggravated~\cite{Inoue2019}. Such loops in a network are found to generate instability in the system dynamics literature~\cite{Senge1980} and more recently in the context of supply chains~\cite{Demirel2019}. In the case of lifting the lockdown in only one prefecture, the loops within that prefecture may magnify its recovery because of the circulation of positive effects in the loops. 

\begin{figure}
    \centering
    \begin{subfigure}[t]{0.495\textwidth}
        \centering
        \includegraphics[width=\linewidth]{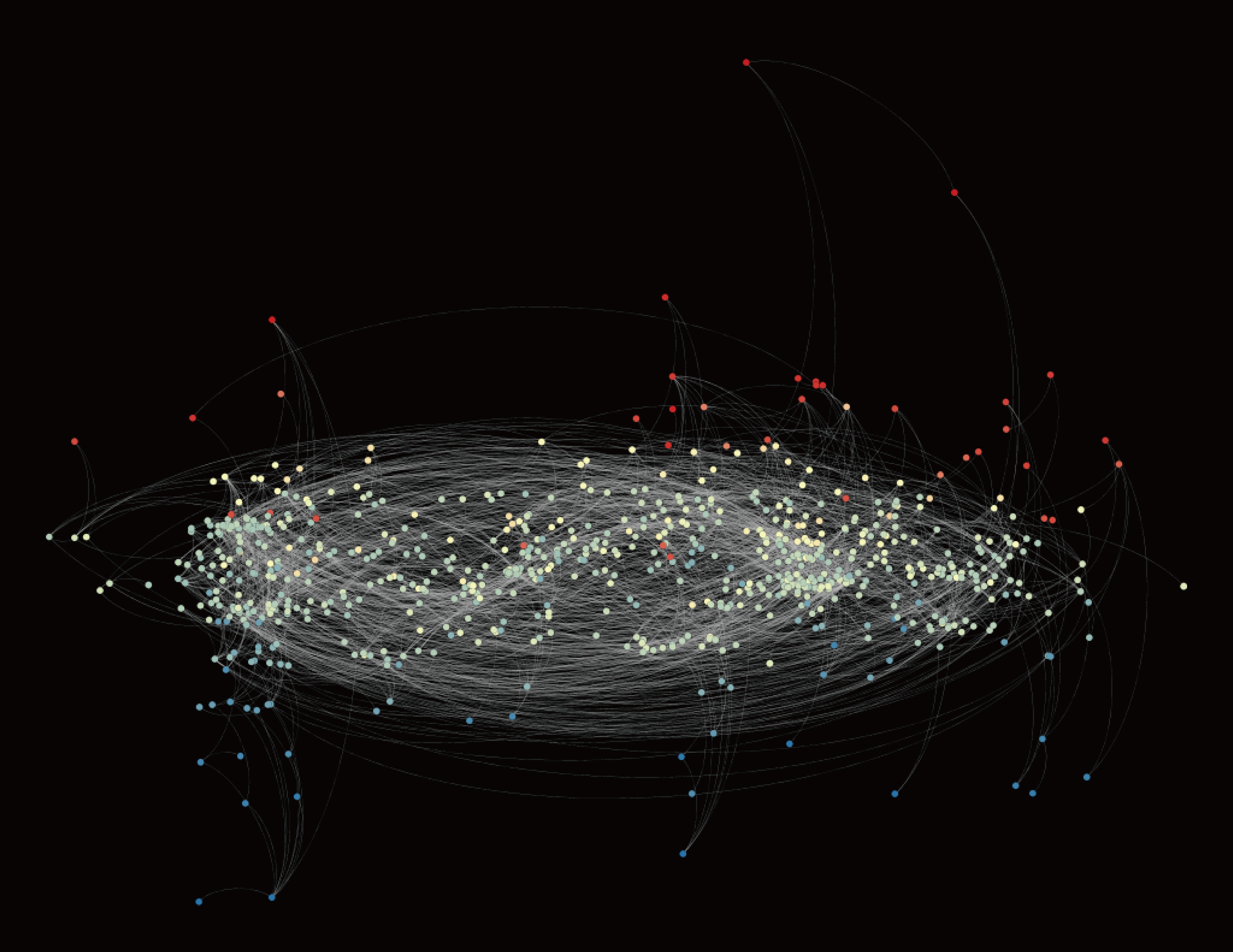} 
    \end{subfigure}
    \hfill
    \begin{subfigure}[t]{0.495\textwidth}
        \centering
        \includegraphics[width=\linewidth]{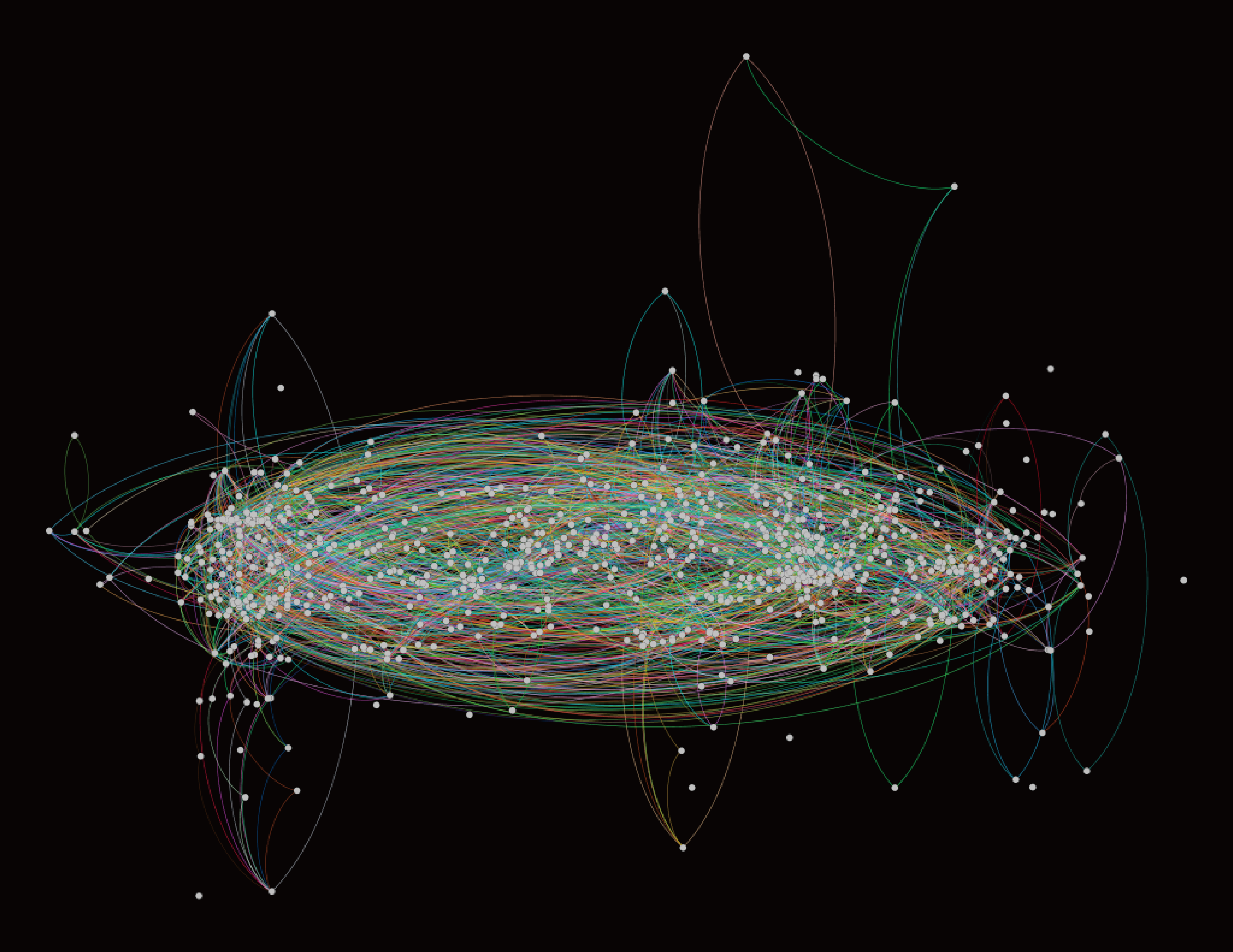} 
    \end{subfigure}
    \caption{Visualisation of supply chains for top 1,000 firms in terms of sales. Each dot indicates a firm. Firms with a higher HH potential are located more upward in both panels. In the left panel, the grey lines illustrate the potential flows computed from the HHD. The red and blue node colours respectively represent higher and lower HH potentials. The right panel shows loop flows computed from HHD, while the different colours represent different cycles.}
    \label{fig:top1000hhd}
\end{figure}

To measure the intensity of the loops in the supply chains within a prefecture, we apply the Helmholtz--Hodge decomposition (HHD) to all the flows in the network and decompose each directed link from firm $i$ to firm $j$, $F_{ij}$, into a potential (or gradient) flow component, $F_{ij}^{(p)}$, and a loop (or circular) flow component, $F_{ij}^{(c)}$~\cite{jiang2011hodge}. Supplementary Information~\ref{ch:hhd} explains the details of the HHD. Figure~\ref{fig:top1000hhd} illustrates potential and loop flows of top 1,000 firms in terms of sales. In particular, the right panel identifies a number of loops in supply chains. Then, our measure of the intensity of the loops for prefecture $a$ is the ratio of the total loop flows within the prefecture $\sum_{i,j \in a} F_{ij}^{(c)}$ to the total degree of all the firms in the prefecture denoted by $F_a$.

Third, we pay attention to the upstreamness of firms in supply chains. Theoretically, upstream firms are affected by supply-chain disruptions through a lack of demand, whereas downstream firms are affected through a lack of supply. However, the effect of upstream and downstream links can differ in size. A recent sectoral analysis~\cite{Branger2019} finds that the profits of more upstream sectors in global value chains are substantially lower than those of more downstream sectors, implying that negative economic shocks propagate upstream more than downstream. To clarify the possible effect of upstreamness, we define the upstream position of each firm $i$ in supply chains by its Helmholtz--Hodge potential, $\phi_i$ computed from the HHD. In other words, the hierarchical position of a firm can be consistently defined by focusing on gradient flows, or all flows less loop flows. The HH potential is higher when the firm is located in a more upstream position. In practice, it is generally higher for firms in the mining, manufacturing, and information and communication sectors, while lower for those in the wholesale, retail, finance, healthcare, and accommodation and food service sectors~\cite{kichikawa2019community}. We average the HH potential over the firms in each prefecture to measure the upstreamness of the prefecture in supply chains (see Supplementary Information~Figure~\ref{fig:hhdmap} for this measure for each prefecture).

Our measure of upstreamness based on the HH potential is conceptually similar to the upstreamness measures developed and widely used in the literature on international trade \cite{Alfaro2019, Antras2012, Antras2013, Fally2012, Fally2017} in that both ours and theirs measure the hierarchical position in supply chains. A stark difference between the two types of measures is that ours is based on firm-level data while others are based on sector-level input-output (IO) tables. Therefore, our measure can incorporate firm-level heterogeneity within the same sector that is ignored in others. In addition, our measure is defined by gradient flows in supply chains that are constructed by eliminating loop flows from all flows. Although many loops at the firm level are found in supply chains, even within the industry \cite{kichikawa2019community}, upstream measures based on IO tables do not incorporate such loops. For these reasons, we will rely on our upstreamness measures at the firm level, not existing measures at the sector level. 

Finally, even when the supplies of parts and components from other prefectures are shut down because of their lockdowns, the negative effect can be mitigated if suppliers can be replaced by those in the prefecture lifting its lockdown. Existing studies~\cite{Barrot2016, Inoue2019} have found that input substitutability can largely mitigate the propagation of negative economic shocks through supply chains. By assumption, suppliers of firms in prefecture $a$ that are in other prefectures on lockdown can be replaced by suppliers in prefecture $a$ that are in the same industry and already connected. To measure the degree of supplier substitutability for prefecture $a$, we divide the number of the latter suppliers by the number of the former.

\paragraph*{Lifting lockdowns in two regions simultaneously} In practice, each prefecture government determined its restriction level of lockdown after observing the spread of COVID-19 in its prefecture and typically ignored the economic interactions with other prefectures through supply chains. This may have led to the misevaluation of the economic effect of lockdown. To emphasise the role of the interactions between prefectures in the economic effect of lockdown, our simulations analyse the economic effect of lifting the lockdown in a prefecture on its GRP when another prefecture lifts its lockdown simultaneously. We define a relative measure of recovery using the ratio of the increase in the GRP of prefecture $a$ when it lifts its lockdown together with prefecture $b$ ($\Delta GRP_a^{ab}$) to its increase when it lifts its lockdown alone ($\Delta GRP_a^{a}$). 

Presumably, the characteristics of the links between the two prefectures largely affect their recovery. Expanding the case of lifting the lockdown in only one prefecture described just above, we are particularly interested in the following variables. First, we define the intensity of the directional links from prefectures $a$ to $b$ and from $b$ to $a$ by
\begin{equation}
\label{eq:link_ab}
    Link_{ab} \equiv \sum_{i \in a, j \in b} F_{ij}/F_a
\end{equation}
and
\begin{equation}
\label{eq:link_ba}
    Link_{ba} \equiv \sum_{i \in a, j \in b} F_{ji}/F_a,
\end{equation}
respectively, where $F_a$ is the total degree of firms in prefecture $a$, as defined before. Second, we focus on potential flows using the HHD as above and define the intensity of potential flows from prefectures $a$ to $b$ and from $b$ to $a$ by 
\begin{equation}
\label{eq:pot_ab}
    Pot_{ab} \equiv \sum_{i \in a, j \in b} F_{ij}^{(p)}/F_a
\end{equation}
and 
\begin{equation}
\label{eq:pot_ba}
    Pot_{ba} \equiv \sum_{i \in a, j \in b} F_{ji}^{(p)}/F_a,
\end{equation}
respectively. Third, the intensity of the loops between prefectures $a$ and $b$ is given by 
\begin{equation}
\label{eq:loop}
    Loop_{ab} \equiv \sum_{i \in a, j \in b} F_{ij}^{(c)}/F_a.
\end{equation}
Supplementary Information~\ref{ch:hhd} describes how to calculate $Pot_{ab}$, $Pot_{ba}$, and $Loop_{ab}$ using a simple example.

Finally, when suppliers of firms in prefecture $a$ are located outside prefectures $a$ and $b$ and thus are locked down, they can be replaced by suppliers in the same industry in prefecture $b$ that are already connected with firms in prefecture $a$. To measure the degree of this supplier substitutability, we divide the total number of the latter suppliers by the total number of the former. 
See Supplementary Information~\ref{ch:subs} for the details.

\section{Results}

\subsection{Simulation of the effect of the actual lockdown}
\label{sec:resultbase}

\begin{figure}[htb]

\centering
\includegraphics[width=.7\linewidth]{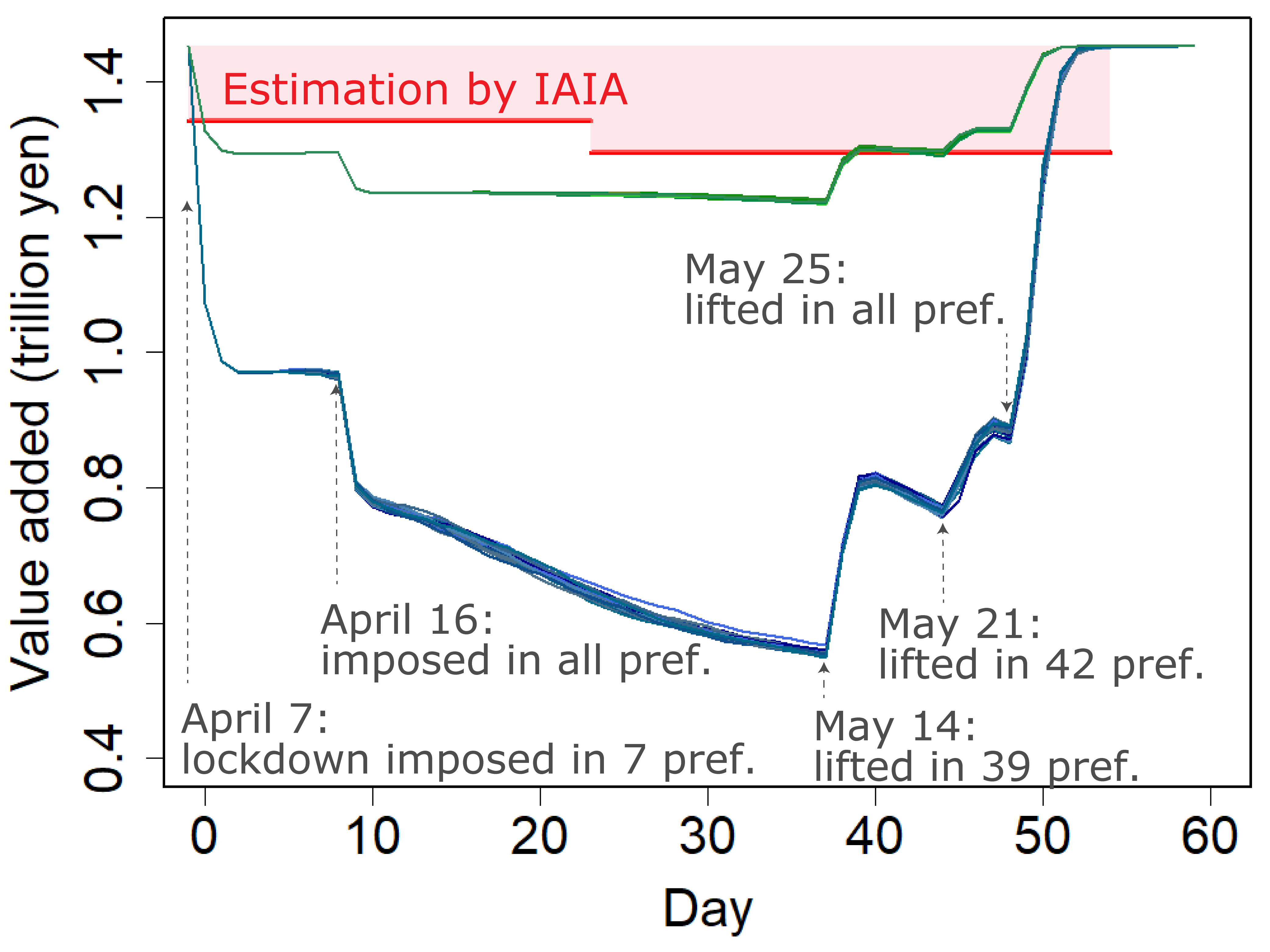}   

\caption{Simulations of value added (GDP) during the actual lockdown. The blue and green lines show the simulation results given the sector-specific rates of reduction in production capacity assumed in the literature \cite{Branger2019, Guan2020} and shown in Supplementary Information~\ref{tbl:mul} and 32.3\% of those rates to calibrate the actual production dynamics, respectively. Each line represents the daily GDP from one Monte Carlo run. The red segments indicate the daily GDP estimated from pre-lockdown GDP and the post-lockdown monthly Indices of All Industry Activity (IAIA) for April and May. }
\label{fig:actual}
\end{figure}

In Figure~\ref{fig:actual}, the blue lines show the results of the 30 Monte Carlo runs conducted to estimate the effect of the actual lockdown in Japan given the sector-specific rates of reduction in production capacity assumed in the literature \cite{Branger2019, Guan2020} and shown in Supplementary Information~\ref{tbl:mul}. The horizontal axis indicates the number of days since the declaration of the state of emergency (April 7) and the vertical axis represents the total value added production, or GDP, of Japan on each day. See Section \ref{sec:soe} for the sequence of the state of emergency across the country. Averaged over the 30 runs, the estimated loss in GDP is 13.1 trillion yen (124 billion U.S. dollars), or 2.5\% of yearly GDP.

\begin{figure}[htb]
\centering

\begin{subfigure}[b]{.42\textwidth}
\includegraphics[width=\linewidth]{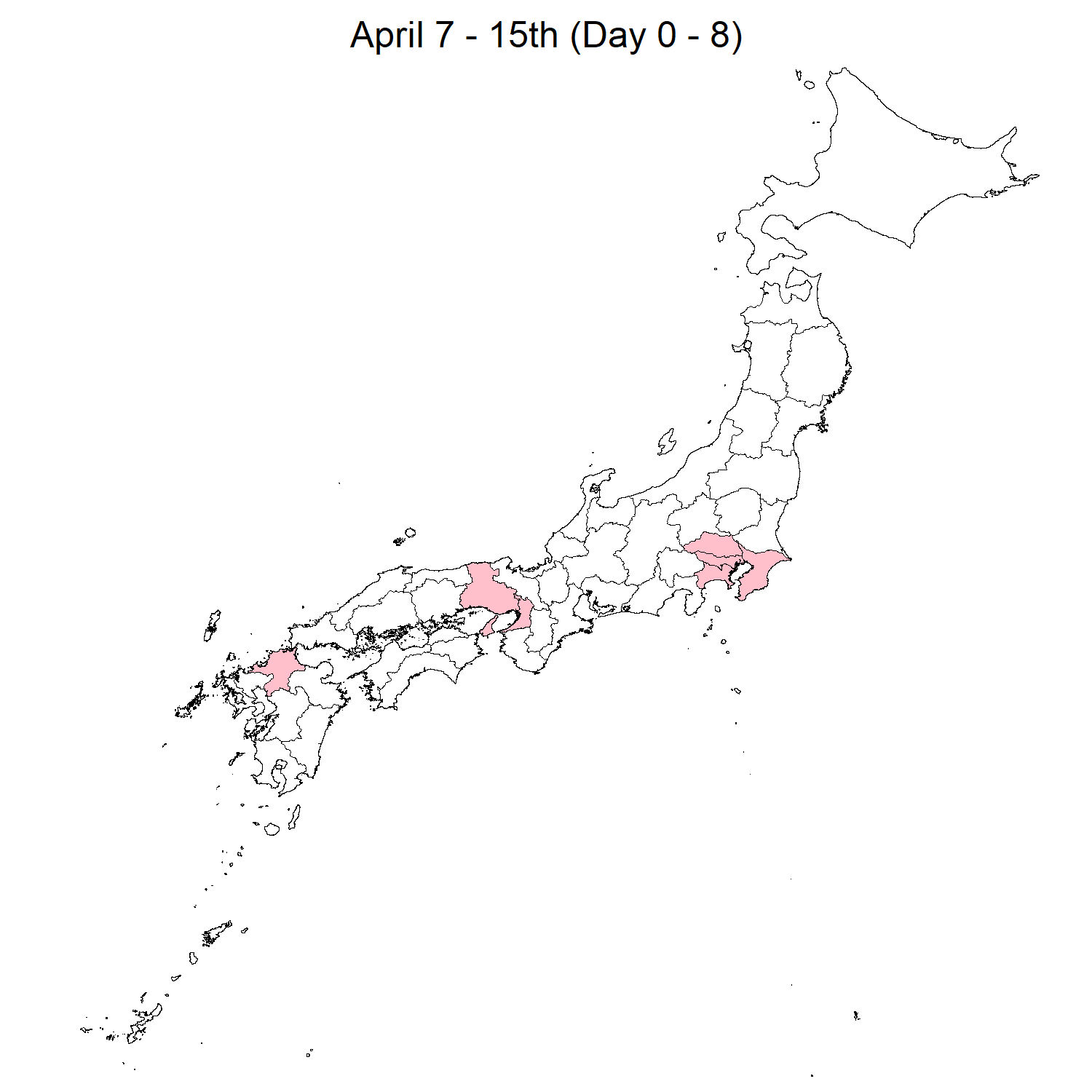}    \end{subfigure} 
\hfill
\begin{subfigure}[b]{.42\textwidth}
\includegraphics[width=\linewidth]{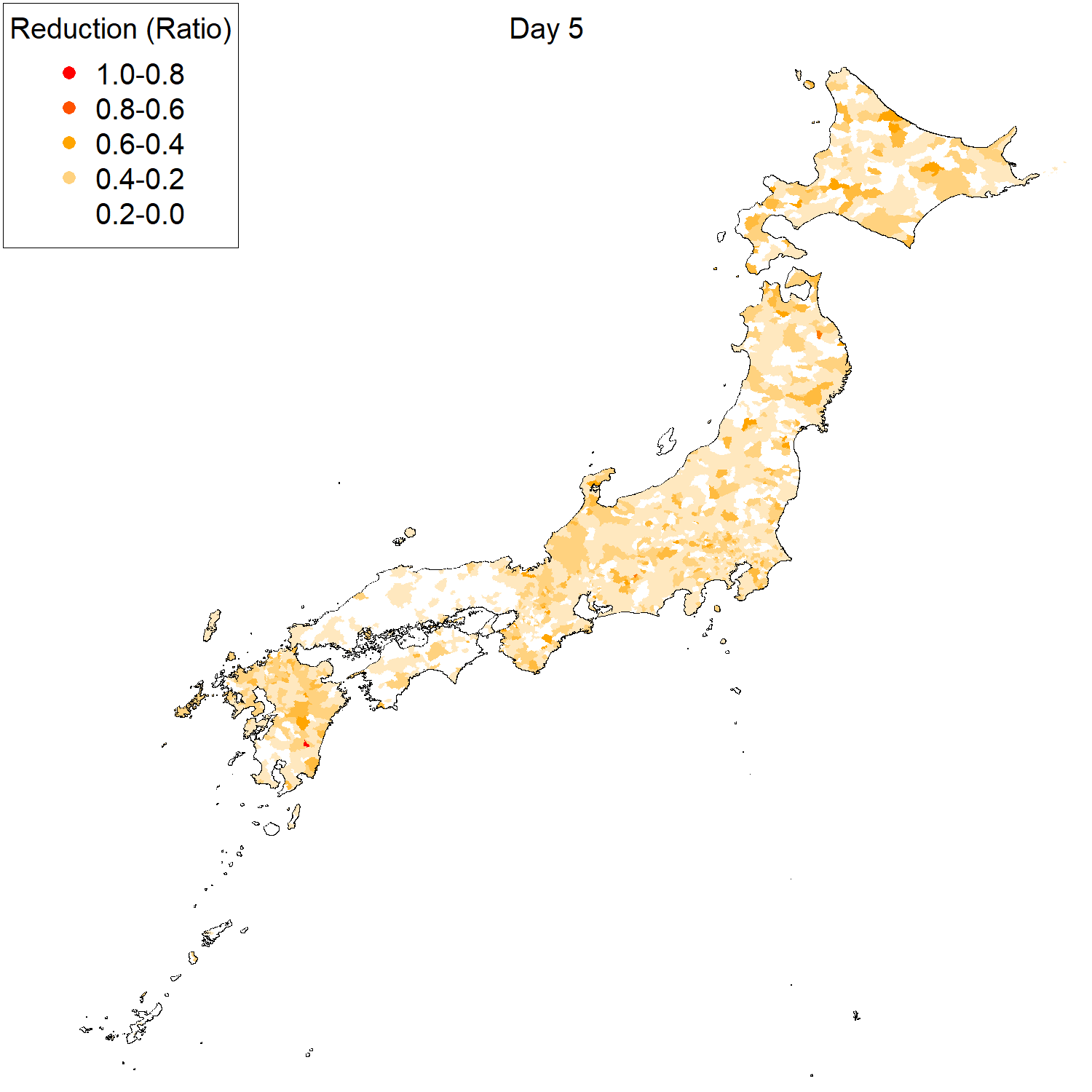}       
\end{subfigure}

\caption{Geographical visualisation of the effect of lockdowns. In the left panel, locked-down prefectures in the first stage of the state of emergency (day 0-8) are shown in red, while the right panel presents the rate of reduction in production averaged over firms in each municipality on day 5, using different colours for different rates of reduction. 
}
\label{fig:actmap}
\end{figure}

Without relying on our model and simulation, we also estimate the changes in daily GDP from pre-lockdown GDP and the post-lockdown monthly Indices of All Industry Activity (IAIA)~\cite{METIIIP2020}. The average daily GDP in April and May estimated from the IAIA is illustrated by the red lines in Figure~\ref{fig:actual} (see Supplementary Information~\ref{ch:Asimu} for the detailed procedures). The total loss of GDP estimated by the IAIA, or the pink area in Figure~\ref{fig:actual}, is 35.0 trillion yen (1.44\% of GDP), 21.5\% of the estimate from our simulations. Our simulation thus overestimates the loss of GDP from the lockdown, possibly because the assumed rates of reduction in production capacity due to the lockdown taken from the literature~\cite{Bonadio2020, Guan2020} are larger than the actual rates in Japan. Therefore, we experiment with different rates of reduction in production capacity by multiplying the benchmark rates by a weight to calibrate changes in production. We find that a weight of 32.3\% results in a close fit between our estimates and those from the IAIA and show the results using green lines in Figure~\ref{fig:actual}.

In either case (blue or green lines), the production loss rises during the lockdown. For example, value added declined monotonically from days 9 to 37 when all prefectures were in the state of emergency, assuming a fixed rate of reduction in production capacity throughout the period. This is because the economic contraction in different regions interacted with each other through supply chains and thus worsened over time. This worsening trend in GDP is consistent with GDP estimated using the IAIA. 

Another notable finding from the simulation is that prefectures that were not locked down were heavily affected by those in lockdowns. To highlight this, the left panel of Figure~\ref{fig:actmap} shows locked-down prefectures in the first stage of the state of emergency (day 0-8) in red, while the lower-right panel presents the rate of reduction in production averaged over firms in each municipality on day 5. From these figures, it is clear that the economic effect of lockdowns in some prefectures diffuse to others that are not locked down. A video is available for the temporal and geographical visualisation. See Appendix~\ref{ch:Asimu}.

In addition, because of the network effect, the earlier lifting of the lockdown in some prefectures does not result in a full recovery of production in these prefectures. Notably, when the lockdown was lifted in 39 prefectures on day 37 (14 May), the simulated GDP showed a sharp recovery but dropped again substantially a few days after the recovery. This drop occurred because the lockdown remained active in eight prefectures including the top two industrial clusters in Japan, namely, greater Tokyo and greater Osaka. Although economic activities returned to normal in these 39 prefectures, their production did not recover monotonically but rather declined again because the major industrial clusters linked with them were still locked down. This finding points to the interactions of the economic effect of lockdown between regions through firm-level supply chains. 

\subsection{Interactions between lockdowns in different regions}
\label{ch:sensitivity}

\begin{figure}[htb]
\centering
\includegraphics[width=0.8\linewidth]{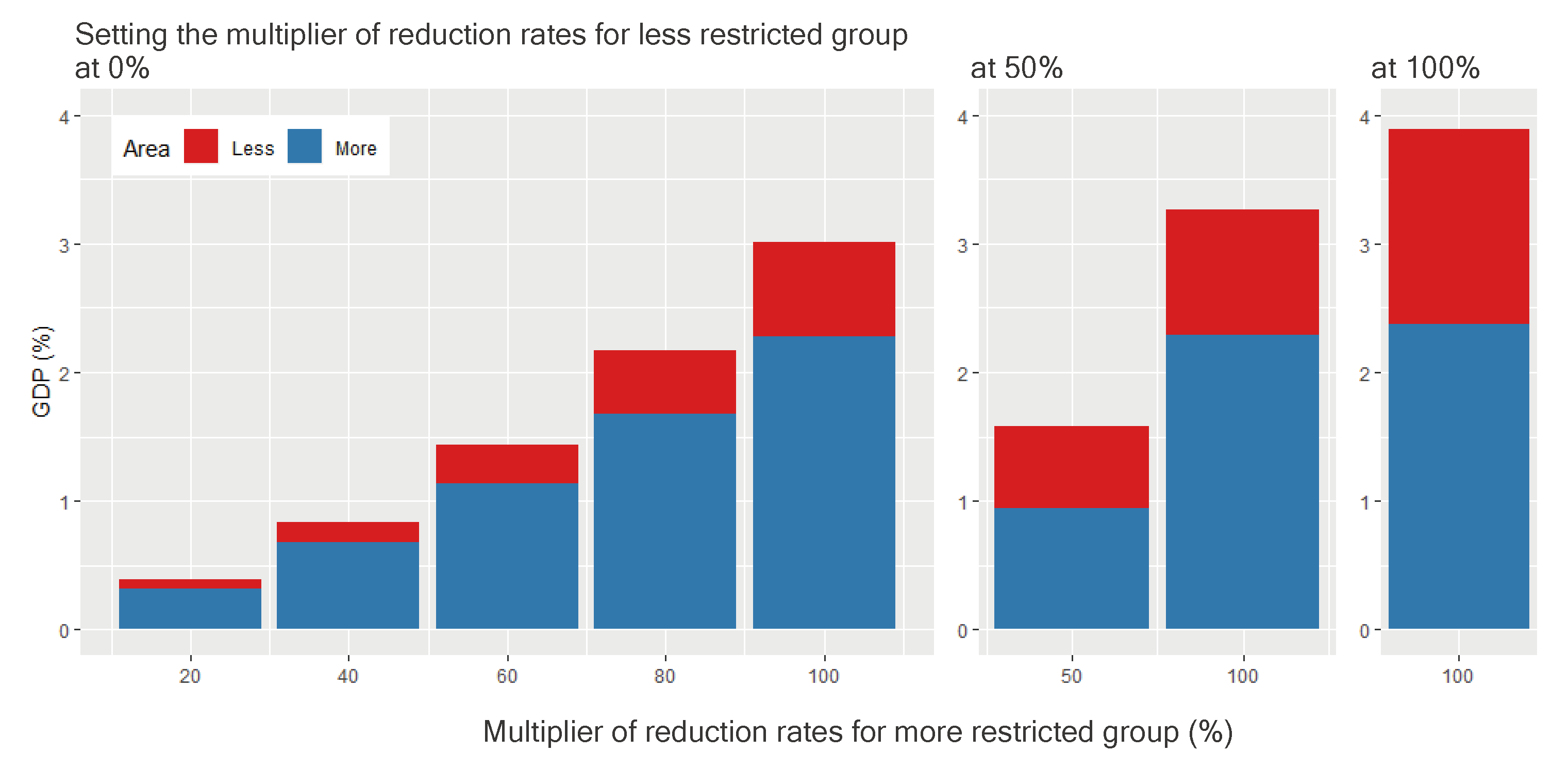}
\caption{Loss in value added as a percentage of total value added (GDP) assuming different restriction levels of lockdown for 60 days between the more and less restricted groups. A restriction level is defined by a multiplier for the sector-specific benchmark rates of reduction in production capacity. For example, the left bar presents the result assuming a multiplier of 0\% (i.e., no restriction) for the less restricted group and 20\% for the more restricted group. The red and blue parts of each bar show the loss of value added in the less and more restricted groups, respectively, as a percentage of GDP.}
\label{fig:stack60}
\end{figure}

Next, we experiment with simulations assuming different restriction levels of lockdown, or different sets of multipliers for the sector-specific benchmark rates of reduction in production capacity, between the more and less restricted groups (Section \ref{sec:methodsim}). The more restricted group comprises the seven prefectures with a large number of COVID-19 cases (pink ones in panel (b) of Figure~\ref{fig:mapsoe}), whereas the less restricted group includes the other 40 prefectures. The left, middle, and right panels of Figure~\ref{fig:stack60} indicate the loss in GDP for different multipliers for the more restricted group when fixing the multiplier for the less restricted group at 0\%, 50\%, and 100\%, respectively. Here, 100\% corresponds to the rates of reduction shown in Supplementary Information~Table~\ref{tbl:mul} and used in the previous subsection and 0\% means no restriction. In each bar, the blue and red parts show the loss of value added in the more and less restricted groups, respectively. 



As shown, the total loss of GDP increases in the restriction level of lockdown in both groups. For example, the total production loss is 1.57\% of GDP when the multiplier is 50\% for both groups (the left bar in the middle panel), while it is larger, or 3.89\%, when the multiplier is 100\% for both (the right panel). More interestingly, the left panel shows that while the less restricted group imposes no restrictions, its value added decreases more (i.e. the red part in Figure~\ref{fig:stack60} increases) as the more restricted group imposes more restrictions. When the level of restrictions in the more restricted group is the highest (i.e. the multiplier is 100\%), the loss in value added in the less restricted group without any lockdown is large: 3.93 trillion yen, or 2.12\% of its pre-lockdown value added. These results clearly indicate that even when prefectures are not locked down, their economies can be damaged because of the propagation of the effect of the lockdowns in other prefectures through supply chains.

\subsection{Effect of lifting the lockdown in one region}
\label{ch:lift}

\begin{figure}[htb]
\centering
\includegraphics[width=.5\linewidth]{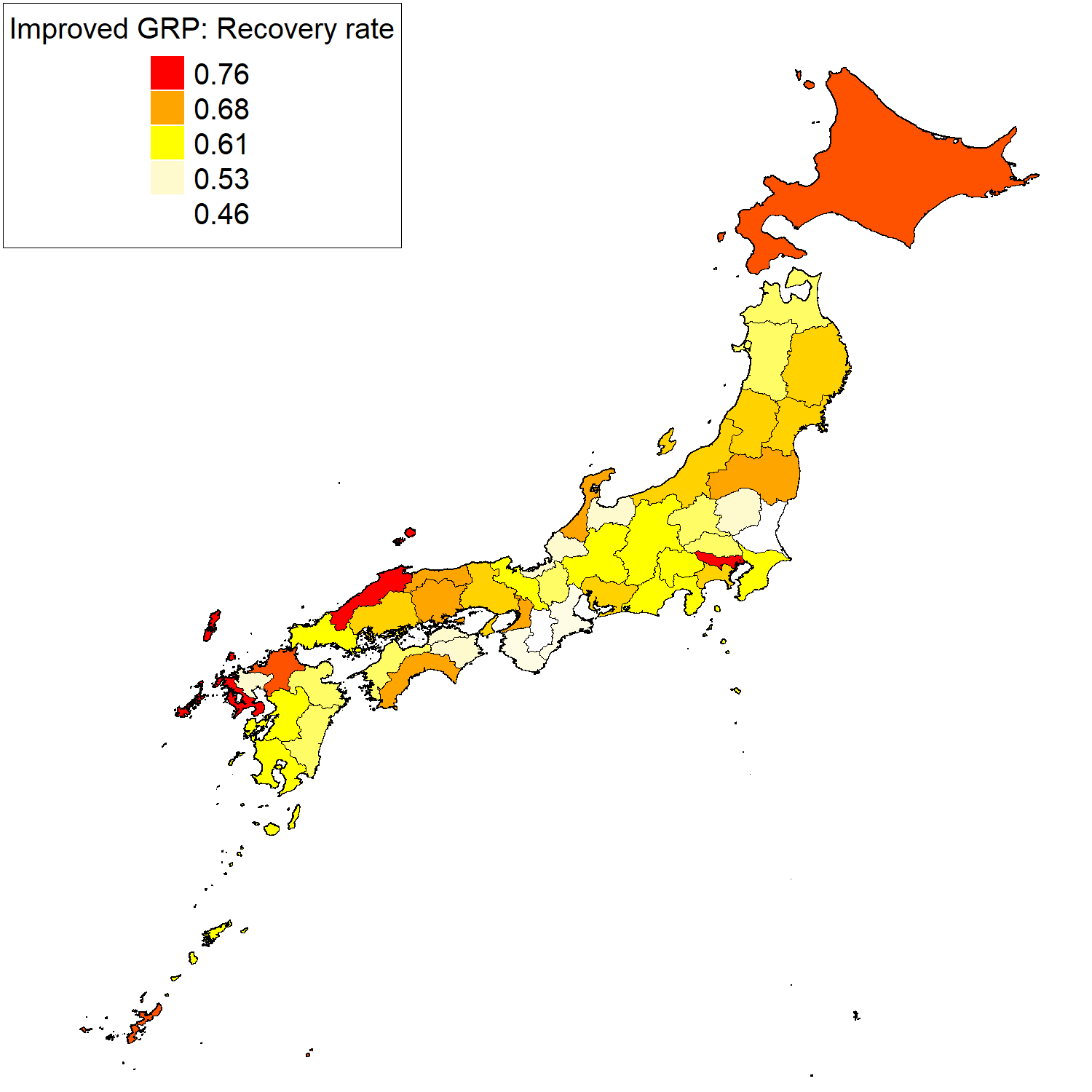}
\caption{Choropleth map of the recovery rate for each prefecture. The recovery rate is defined as the ratio of the total gain of a prefecture's GRP from lifting its own lockdown to its total loss from the lockdown of all the prefectures for two weeks. 
}
\label{fig:hmap}
\end{figure}

We further examine, when only one prefecture lifts its lockdown, while all the other prefectures remain locked down, how the recovery of that prefecture from lifting its lockdown is determined by its network characteristics. Figure~\ref{fig:hmap} illustrates the recovery rate of each prefecture defined as the ratio of the total gain of its value added, or gross regional production (GRP), from lifting the lockdown to its total loss from the lockdown of all the prefectures for two weeks. Red prefectures recover the most, yellow ones recover moderately, and white ones recover slightly. See Supplementary Information~Figure~\ref{fig:onoffeach} for the recovery rate of each prefecture. 

One notable finding from this figure is that the prefectures that recover the most, or the red prefectures in Figure~\ref{fig:hmap}, include Hokkaido, Shimane, and Okinawa, which are remote from industrial hubs in terms of both geography and supply chains, suggesting the effect of network characteristics on economic recovery by lifting a lockdown (see Supplementary Information~Figure~\ref{fig:linkMap} for inter-prefecture supply chains and Supplementary Information~Figure~\ref{fig:prefInfo} for the name and location of each prefecture). 

\begin{figure}[htb]
\centering
\includegraphics[width=.80\linewidth]{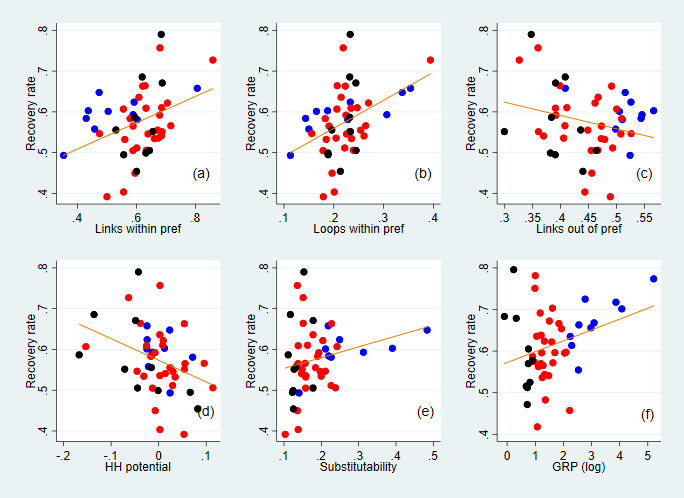}
\caption{Correlation between the recovery rate and selected network measures. The vertical axis indicates the recovery rate defined as the ratio of the increase in the GRP of a prefecture by lifting its own lockdown to its decrease because of the lockdown of all the prefectures. Except for panel (f), the effect of GRP is excluded from the recovery rate. The horizontal axis indicates 
the share of the links within the prefecture to its all links in (a), 
the share of the loop flows within the prefecture to its total flows in (b), 
the share of the links to other prefectures to all links in (c), 
the standardised potential flows in (d), 
the share of substitutable suppliers to all suppliers outside the prefecture in (e), and GRP in logs in panel (f), 
The orange line in each panel specifies the fitted value from a linear regression that controls for the effect of GRP.
The blue, black, and red dots show prefectures whose GRP is among the top 10, bottom 10, and others, respectively.}
\label{fig:OneLift_comb}
\end{figure}

We further examine the correlation between the recovery rate and network measures explained in Section \ref{sec:methodsim} (i.e. those for isolation, loops, upstreamness, and supplier substitution) and test the significance of the correlation using ordinary least squares (OLS) estimations. Figure~\ref{fig:OneLift_comb} illustrates the correlation between the recovery rate and network measures. To control for the effect of the prefecture's economic size on its recovery (Figure~\ref{fig:OneLift_comb}(f)), we include GRP in logs in all the OLS estimations and exclude the effect of GRP from the recovery rate in Figure~\ref{fig:OneLift_comb}. The number of links of each prefecture could also be controlled for; however, because its correlation coefficient with GRP is 0.965 (Supplementary Information~Table~\ref{tbl:OneLift_corr}), we do not use the total links in our regressions to avoid multicollinearity. Supplementary Information~Table~\ref{tbl:OneLift_reg} presents the OLS results. 

In panels (a) and (b) of Figure~\ref{fig:OneLift_comb}, the supply-chain links and loops within the prefecture are found to be positively correlated with the recovery rate. These results suggest that when a prefecture is more isolated in the network and has more loops within it, the positive effect of lifting a lockdown circulates in the loops, which can mitigate the propagation of the negative effects of other prefectures' lockdowns. By contrast, the outward links to other prefectures and the HH potential of the prefecture are negatively and significantly correlated with the recovery rate (panels (d) and (e)). These findings imply that prefectures with more upstream firms in supply chains tend to recover less from lifting their own lockdowns. Panel (f) indicates that the recovery rate is higher when more suppliers in other prefectures under lockdown can be replaced by those in the prefecture lifting its lockdown.

\subsection{Effect of lifting the lockdowns in two regions simultaneously}
\label{ch:lifttwo}

\begin{figure}[htb]
\centering
\includegraphics[width=.95\linewidth]{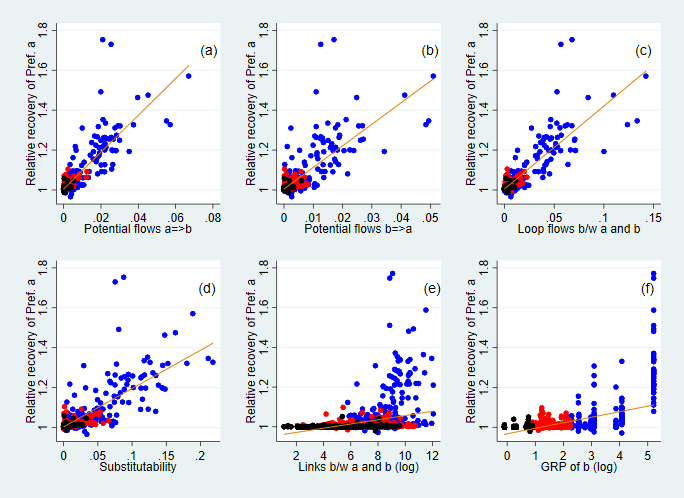}
\caption{Correlation between the relative recovery and selected network measures. The vertical axis indicates the relative recovery of prefecture $a$, defined as the ratio of the increase in the GRP of prefecture $a$ by lifting its lockdown together with prefecture $b$ to its increase by lifting its lockdown alone. The effect of the GRP of $b$ and total links between the two are excluded from the relative recovery measure. The variable in the horizontal axis is given by 
Equations \ref{eq:pot_ab} and \ref{eq:pot_ba} in panels (a) and (b), respectively, 
Equation \ref{eq:loop} in (c), 
the share of substitutable suppliers in $b$ for those in $a$ among $a$'s locked-down suppliers in (d),
the number of links between prefectures $a$ and $b$ in (e)
and the GRP of $b$ in logs in (f). 
The orange line in each panel signifies the fitted value from a linear regression that controls for the effect of the GRP of $b$ and total number of links between $a$ and $b$ in (a)--(d).
The blue, black, and red dots show the pairs of prefectures $a$ and $b$ for which the GRP of $b$ is among the top 10, bottom 10, and others, respectively.}
\label{fig:DoubleLift_comb}
\end{figure}

Finally, we simulate the effect on the production of prefecture $a$ if it lifted its lockdown together with prefecture $b$. We compare the recovery in prefecture $a$'s GRP by lifting its lockdown together with prefecture $b$ and that by lifting its lockdown alone and compute the relative recovery measure, as shown in Supplementary Information~Figure~\ref{fig:DoubleLift_prefcode}. Using a regression framework as above, we investigate how the relative recovery measure of prefecture $a$ is affected by the network relationships between prefectures $a$ and $b$. Figure~\ref{fig:DoubleLift_comb} illustrates the correlation between selected key variables and the relative recovery. In the regression analysis, we always control for the GRP of prefecture $b$, its squares, and the number of links between and prefectures $a$ and $b$ that may affect the relative recovery (Figure~\ref{fig:DoubleLift_comb}~(e) and (f)). Then, we exclude these effects from the relative recovery in panels (a)--(d) in the figure. Supplementary Information~Table~\ref{tbl:DoubleLift_reg} presents the results of the OLS estimations.

Panels (a) and (b) of Figure~\ref{fig:DoubleLift_comb} show that even after controlling for the effect of economic size and number of links between the two prefectures, the ratio of potential flows from prefecture $a$ to $b$ and from $b$ to $a$ to the total flows of $a$ is positively correlated with the relative recovery. Supplementary Information~Figure~\ref{fig:DoubleLift_comb2} shows similarly positive correlation between for the number of links between the two, rather than potential flows, and the recovery. These results suggest that the recovery from lifting a lockdown is greater when two prefectures closely linked through their supply chains, regardless of the direction, lift their lockdowns together. Further, we find that prefecture $a$ recovers more when prefectures $a$ and $b$ are linked through more circular flows (panel (c)), confirming that the positive impacts of lifting a lockdown can circulate and be strengthened in inter-regional supply-chain loops. Panel (d) indicates that if prefecture $a$'s suppliers in other prefectures are locked down but can be replaced by suppliers in prefecture $b$ easily, prefecture $a$'s recovery is higher when the two prefectures lift their lockdowns together. 
Although the correlation between the relative recovery measure and network variables seems to be largely driven by the observations for which the GRP of prefecture $b$ is large (depicted by the blue dots in Figure~\ref{fig:DoubleLift_comb}), we find that the positive correlation still exists without these observations (Supplementary Information~Figure~\ref{fig:DoubleLift_comb_notop10}).

\section{Discussion and Conclusion}

Our simulation analysis reveals that the economic effects of lockdowns in different regions interact with each other through supply chains. Our results and their implications can be summarised as follows. 

First, when a firm is locked down, its suppliers and customer firms are affected because of a lack of demand and supply, respectively. Therefore, the production of regions can recover more by lifting their lockdowns together when they are closely linked through supply chains in either direction (Figure~\ref{fig:DoubleLift_comb}(a)--(b)). Besides the total number of links between the two regions, the intensity of such links compared with those with others is also important.

Second, when the firms in a region are in more upstream positions in the whole network or they are mostly suppliers of simple parts, the production of the region does not recover substantially by lifting its lockdown alone (Figure~\ref{fig:OneLift_comb}(d)). Although the negative economic effect of lockdown can propagate downstream and upstream, firms can mitigate downstream propagation easily using inventory or replacing locked-down suppliers. The difference between the downstream and upstream effects of lockdown is aggravated as the effect propagates further through supply chains. This finding is in line with the literature~\cite{Branger2019, Fu2018} that also finds the upstream accumulation of negative effects on profits and assets. In practice, our result implies that a region with many small- and medium-sized suppliers of simple materials and parts should be cautious about whether it lifts its lockdown, which may not result in a large economic benefit but still promote the spread of COVID-19. 

Third, the production of a region can recover more by lifting its lockdown when it is more isolated in the network or embodies more supply-chain loops within the region (Figures \ref{fig:OneLift_comb}(a) and (b)). Similarly, the production of the two regions can recover more by lifting their lockdowns together when their inter-regional links have more loops (Figure~\ref{fig:DoubleLift_comb}(c)). These results imply that the positive economic effect of lifting a lockdown circulates and is intensified in loops, consistent with those in~\cite{Inoue2019}. Supply-chain loops exist between two regions when the final goods produced are used as inputs by suppliers, while suppliers provide parts and components to final-good producers and the loop stretches across two regions. The importance of loops in the diffusion of the economic effects in networks is not fully recognised either in the academic literature or in policymaking. 

Finally, the recovery of a region from its lockdown is greater when suppliers still locked down can be replaced by those within the region or in other regions without a lockdown in place (Figures \ref{fig:OneLift_comb}(e) and \ref{fig:DoubleLift_comb}(f)). The role of the substitutability of suppliers in mitigating the propagation effect through supply chains has been empirically found in the literature~\cite{Barrot2016, Kashiwagi2018, Inoue2019, Inoue2019b}. In practice, this finding suggests two management strategies for regional governments and firms. To minimise the economic loss from lockdown, a region should develop a full set of industries to allow the substitution of any industry to be possible. Alternatively, the firms in the region should be linked with geographically diverse suppliers so that suppliers in a locked-down region can be replaced by those in other regions without a lockdown.

All these results point to the need for policy coordination among regions when regional governments impose or lift a lockdown. Although this study uses the inter-firm supply chains within a country and considers the economic effect of prefecture-level lockdowns, our results can be applied to the effect of country-level lockdowns propagating through international supply chains. For example, many suppliers of German firms are located in Eastern Europe and many suppliers of US firms are in Mexico. Our results thus suggest that the economic gains of Eastern Europe and Mexico from lifting their lockdowns is minimal if Germany and the United States, respectively, remain locked down. In addition, our framework can be applied to the case of other infectious diseases, and it is likely to suggest a need for the inter-regional and international coordination of lockdown strategies to prevent the spread of infection. 


\section*{Acknowledgement}
This research used the computational resources of the supercomputer Fugaku (the evaluation environment in the trial phase) provided by the RIKEN Center for Computational Science. OACIS~\cite{murase2017open} and CARAVAN~\cite{murase2018caravan} were used for the simulations in this study.
This research was conducted as part of a project entitled `Research on relationships between economic and social networks and globalization' undertaken at the Research Institute of Economy, Trade, and Industry (RIETI). 
The authors are grateful for the financial support of JSPS Kakenhi Grant Nos. JP18K04615 and JP18H03642. 
We thank Yoshi Fujiwara for advising how to calculate the HHD.

\section{Data Availability}
The data that support the findings of this study are available from Tokyo Shoko Research (TSR), but restrictions apply to the availability of these data, which were used under license for the current study, and so are not publicly available. Data are however available with permission of Tokyo Shoko Research (TSR).

\bibliographystyle{unsrt}
\bibliography{main}

\newpage

\appendix

\noindent{\huge {\bf Supplementary Information}}

\renewcommand\thefigure{\thesection.\arabic{figure}} 
\renewcommand\thetable{\thesection.\arabic{table}} 

\section{Data}
\label{ch:appdata}
\setcounter{figure}{0} 
\setcounter{table}{0} 

\subsection{Supply chains}

In the TSR data, the maximum number of suppliers and customers reported by each firm is 24. However, we can capture more than 24 by looking at supplier--customer relations from the opposite direction. Because the TSR data include the address of the headquarters of each firm, we can identify the longitude and latitude of each headquarters using the geocoding service provided by the Center for Spatial Information Science at the University of Tokyo.

Because the TSR data do not include the value of each transaction between two firms, we estimate it in two steps. First, we divide each supplier's sales into its customers in proportion to the sales of customers to obtain a tentative sales value. Second, we employ 2015 IO Tables for Japan~\cite{METI15} to transform these tentative values into more realistic ones. Specifically, we aggregate the tentative values at the firm-pair level to obtain the total sales for each pair of sectors. We then divide the total sales for each sector pair by the transaction values for the corresponding pair in the IO tables. The ratio is then used to estimate the transaction values between firms. The final consumption of each sector is allocated to all the firms in the sector using their sales as weights. 

\begin{figure}[htb]
\centering
\includegraphics[width=0.6\linewidth]{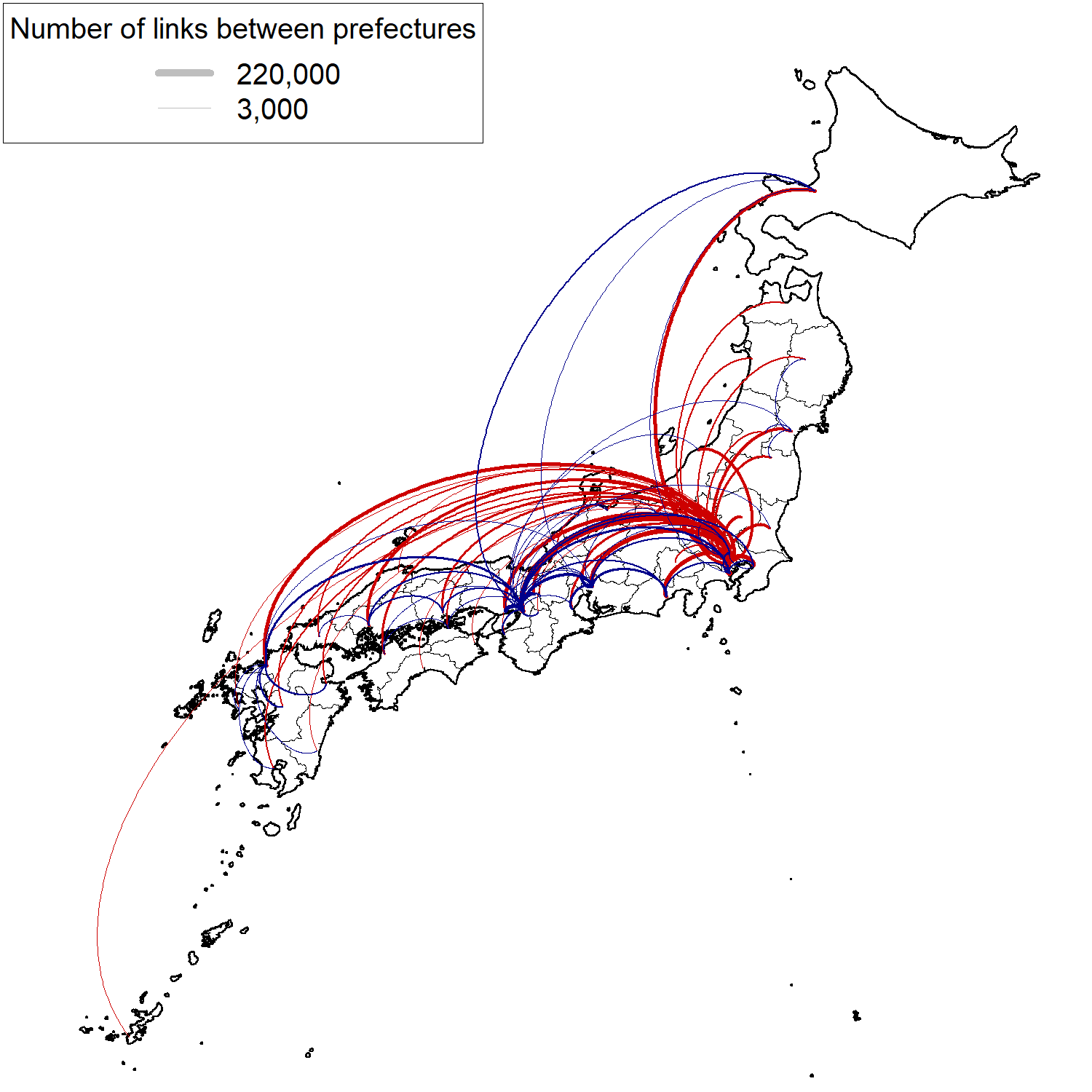} 
\caption{Inter-prefectural links. Inter-firm links are aggregated into inter-prefectural links, ignoring the directions of the links. The inter-prefectural links between two prefectures are not shown here if the number of inter-firm links is less than 3,000. The links within each prefecture are also ignored. The red and blue lines show the inter-prefectural links between Tokyo and other prefectures and between two of other prefectures, respectively.}
\label{fig:linkMap}
\end{figure}

Although the supply chains used in our simulations are at the firm level, this study often uses features of the supply chains at the prefecture level because different prefectures imposed lockdowns to different degrees. Therefore, Figure~\ref{fig:linkMap} illustrates the inter-prefecture supply chains.
The red and blue lines show the inter-prefectural links between Tokyo and other prefectures and between other prefectures than Tokyo, respectively.
We observe that Tokyo is the centre of supply chains in Japan, while several smaller hubs such as Aichi, Osaka, and Fukuoka also exist.

\subsection{Prefectures in Japan}
\label{ch:prefmap}

\begin{figure}[htb]
\centering
\includegraphics[width=0.8\linewidth]{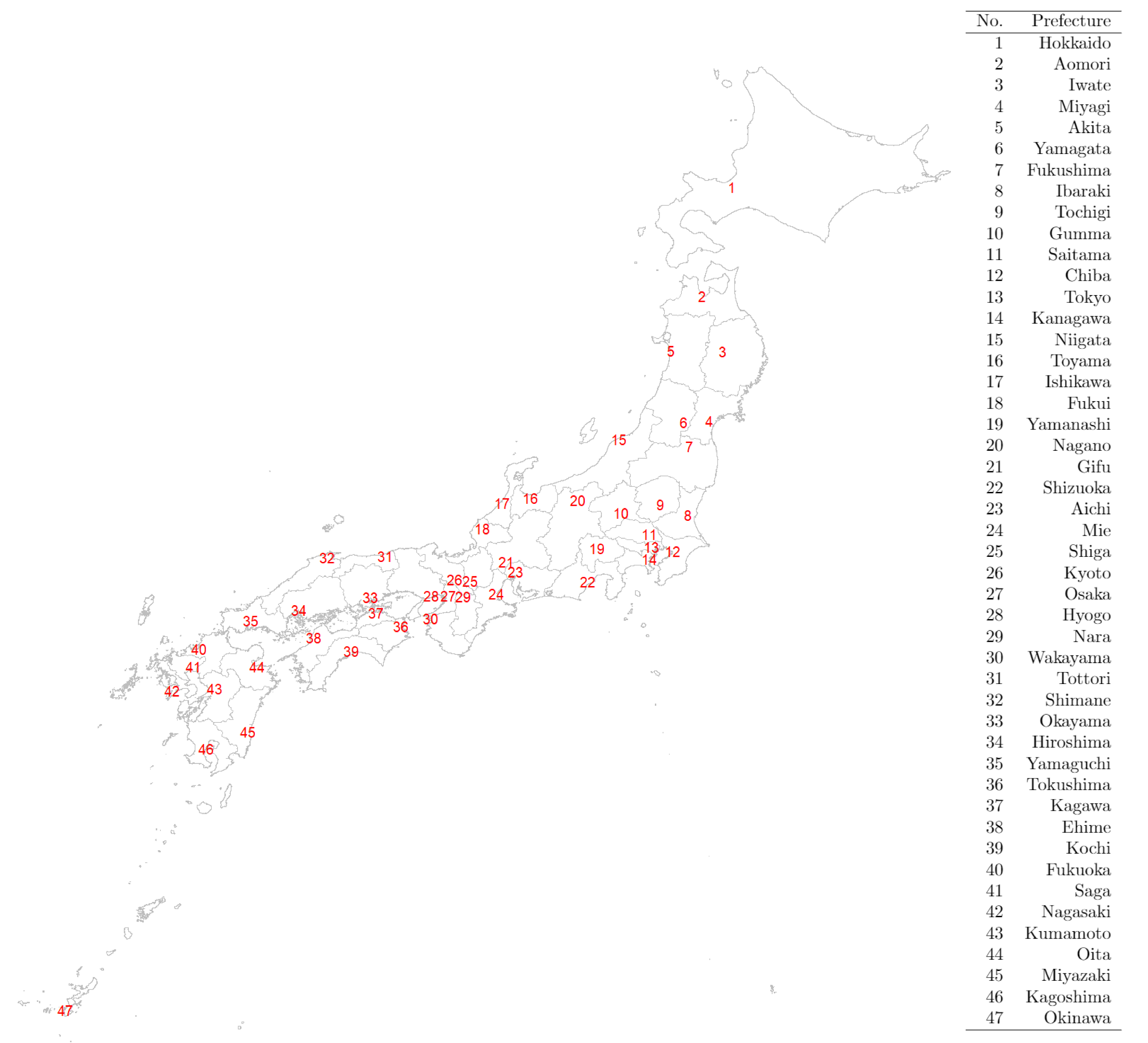}
\caption{Prefecture locations and their codes. The number on the map is the JIS code of each prefecture shown in the right table. 
}
\label{fig:prefInfo}
\end{figure}

Because this study uses prefectures as the unit of regions, it is important to provide information on the prefectures in Japan.
Figure~\ref{fig:prefInfo} shows the locations, Japan Industrial Standard (JIS) codes, and names of the 47 prefectures. In Figures~\ref{fig:onoffoverall} and
the JIS codes are shown on the horizontal axis.

\newpage
\clearpage

\subsection{Geographic presentation of the timeline of lockdowns}
\label{ch:lockmap}

\begin{figure}[b]
    \centering
    \begin{subfigure}[t]{0.3\textwidth}
        \centering
        \includegraphics[width=\linewidth]{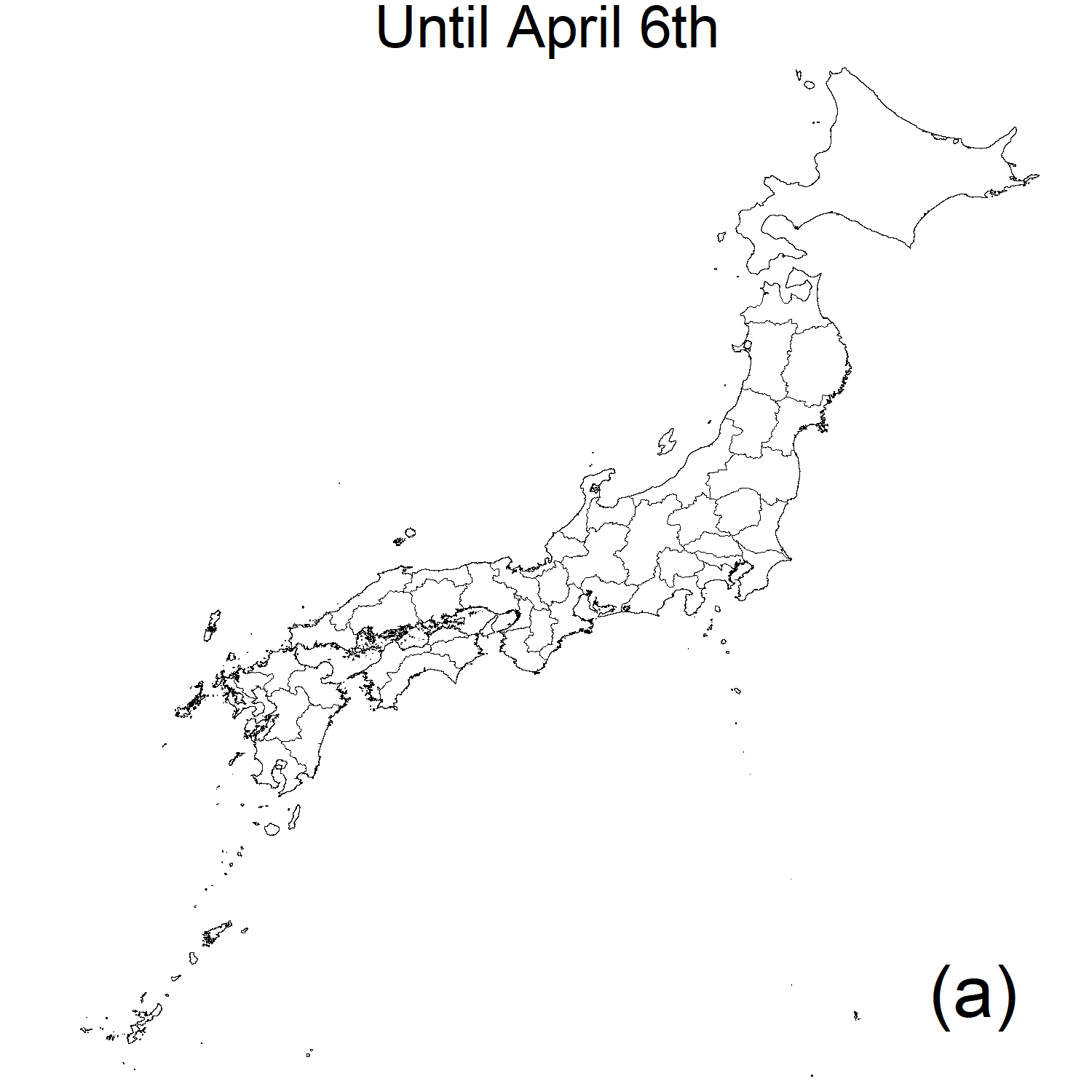} 
    \end{subfigure}
    \hfill
    \begin{subfigure}[t]{0.3\textwidth}
        \centering
        \includegraphics[width=\linewidth]{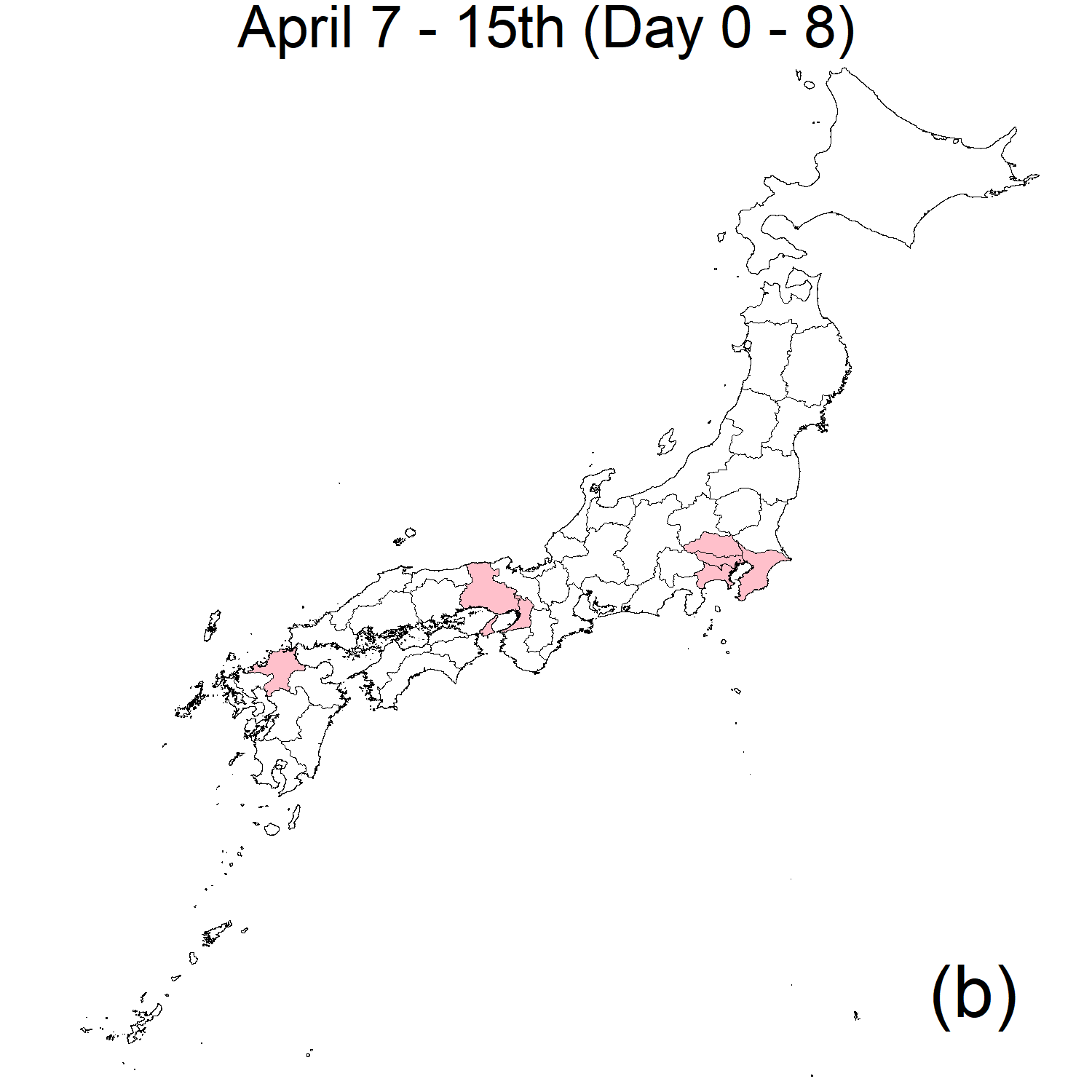} 
    \end{subfigure}
    \hfill
    \begin{subfigure}[t]{0.3\textwidth}
        \centering
        \includegraphics[width=\linewidth]{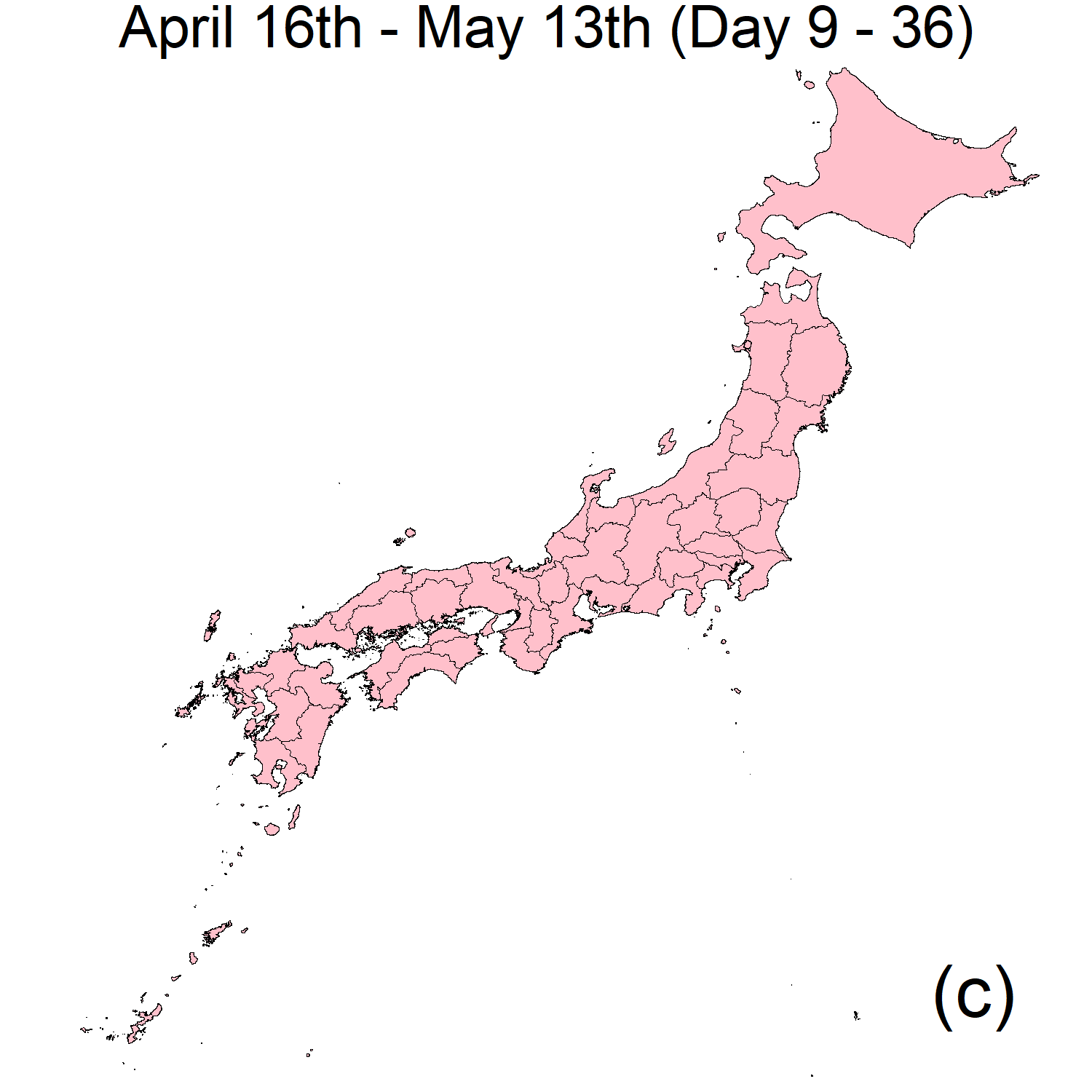} 
    \end{subfigure}

    \vspace{1ex}

    \begin{subfigure}[t]{0.3\textwidth}
        \centering
        \includegraphics[width=\linewidth]{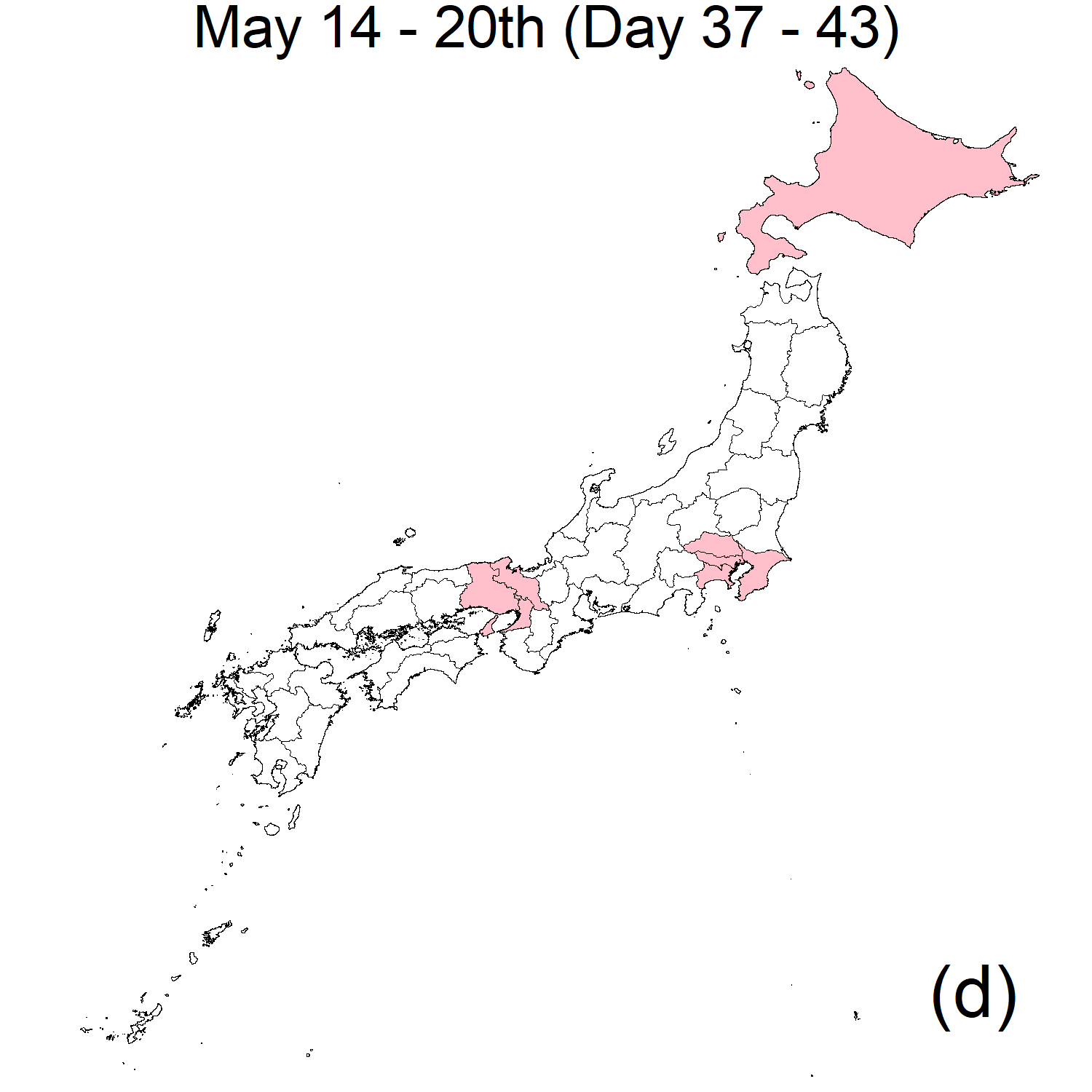} 
    \end{subfigure}
    \hfill
    \begin{subfigure}[t]{0.3\textwidth}
        \centering
        \includegraphics[width=\linewidth]{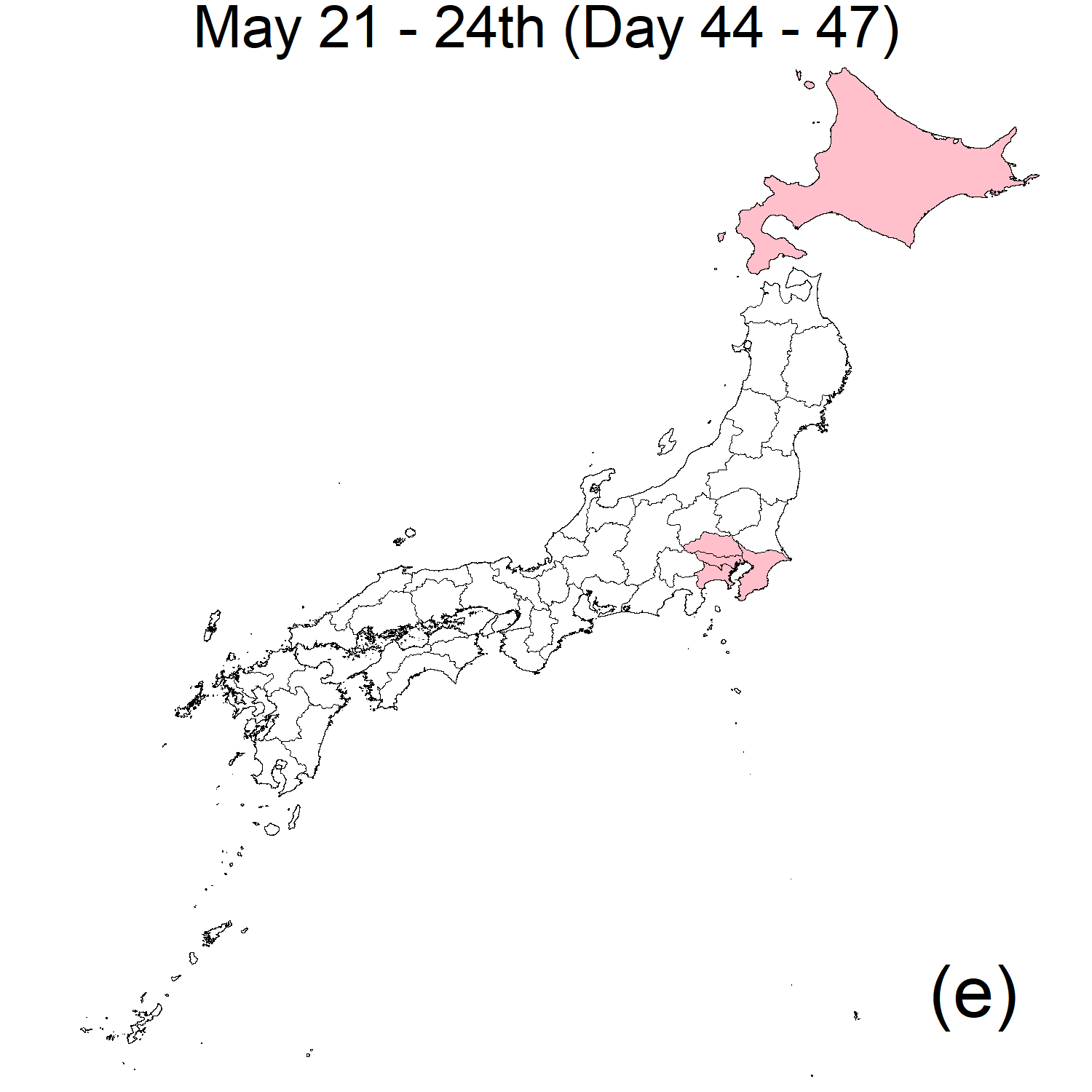} 
    \end{subfigure}
    \hfill
    \begin{subfigure}[t]{0.3\textwidth}
        \centering
        \includegraphics[width=\linewidth]{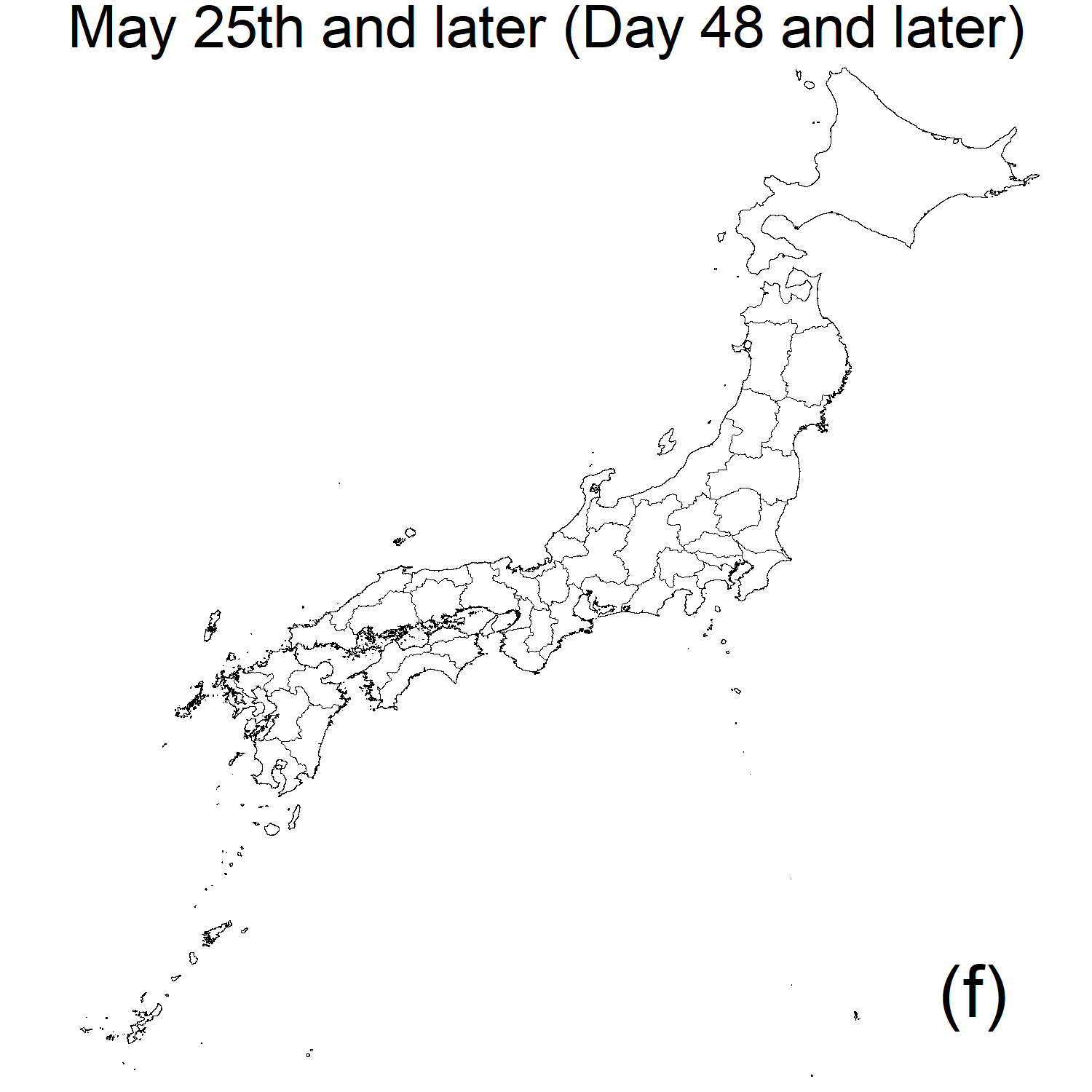} 
    \end{subfigure}

    \caption{Changes in locked down prefectures. The pink prefectures in each panel show those that were locked down during the period.}
    \label{fig:mapsoe}
\end{figure}

Supplementary Information~Figure~\ref{fig:mapsoe} shows where and when the lockdowns were imposed to prefectures.

\section{Methods}
\subsection{Model}
\label{ch:appmodel}
\setcounter{figure}{0} 
\setcounter{table}{0} 

We rely on the model of Inoue and Todo~\cite{Inoue2019, Inoue2019b}, an extension of the existing agent-based models used to examine the propagation of shocks by natural disasters through supply chains, including Hallegatte's model~\cite{Hallegatte08}. Each firm uses a variety of intermediates as inputs and delivers a sector-specific product to other firms and final consumers. Firms have an inventory of intermediates to address possible supply shortages. 

In the initial stage before an economic shock, the daily trade volume from supplier $j$ to customer $i$ is denoted by $A_{i,j}$, whereas the daily trade volume from firm $i$ to final consumers is denoted by $C_i$. Then, the initial production of firm $i$ in a day is given by 
\begin{equation}
P_{\mbox{ini}i}=\Sigma_j{A_{j,i}}+C_i.
\label{eq:p}
\end{equation}
On day $t$ after the initial stage, the previous day's demand for firm $i$'s product is $D_i^* (t-1)$. The firm thus makes orders to each supplier $j$
so that the amount of its product of supplier $j$ can meet this demand, $A_{i,j}{D_i^*(t-1)}/{P_{\mbox{ini}i}}$. We assume that firm $i$ has an inventory of the intermediate goods produced by firm $j$ on day $t$, $S_{i,j} (t)$, and aims to restore this inventory to a level equal to a given number of days $n_i$ of the utilisation of the product of supplier $j$. The constant $n_i$ is assumed to be Poisson distributed, where its mean is $n$, which is a parameter. In addition, $n_i$ does not take a number smaller than 4, although the model in the previous literature sets this number to 2. Since the small minimum inventory size causes a bullwhip effect (fluctuation of production level), we set the number to 4 in this work and recalibrate the parameters. When the actual inventory is smaller than its target, firm $i$ increases its inventory gradually by $1/\tau$ of the gap, so that it reaches the target in $\tau$ days, where $\tau$ is assumed to be 6 to follow the original model~\cite{Hallegatte08}. Therefore, the order from firm $i$ to its supplier $j$ on day $t$, denoted by $O_{i,j}(t)$, is given by 
\begin{equation}
O_{i,j}(t)=A_{i,j}\frac{D_i^*(t-1)}{P_{\mbox{ini}i}}+\frac{1}{\tau}\left[n_i A_{i,j}-S_{i,j}(t)\right],
\label{eq:o}
\end{equation}
where the inventory gap is in brackets. Accordingly, total demand for the product of supplier $i$ on day $t$, $D_i(t)$, is given by the sum of final demand from final consumers and total orders from customers:
\begin{equation}
D_i(t)=\Sigma_jO_{j,i}(t)+C_i.
\end{equation}

Now, suppose that an economic shock hits the economy on day 0 and that firm $i$ is directly affected. Subsequently, the proportion $\delta_{i}(t)$ of the production capital of firm $i$ is malfunctioning. In this study, $\delta_{i}$ is determined by the sector and prefecture to which firm $i$ belongs, and a term for which a lockdown is imposed. Hence, the production capacity of firm $i$, defined as its maximum production assuming no supply shortages, $P_{\mbox{cap}i}(t)$, is given by
\begin{equation}
P_{\mbox{cap}i}(t)=P_{\mbox{ini}i}(1-\delta_i(t)).
\end{equation}
The production of firm $i$ might also be limited by the shortage of supplies. Because we assume that firms in the same sector produce the same product, the shortage of supplies suffered by firm $j$ in sector $s$ can be compensated for by supplies from firm $k$ in the same sector $s$. Firms cannot substitute new suppliers for affected suppliers after the disaster, as we assume fixed supply chains. Thus, the total inventory of the products delivered by firms in sector $s$ in firm $i$ on day $t$ is
\begin{equation}
S_{\mbox{tot}i,s}(t)=\Sigma_{j\in s}S_{i,j}(t).
\end{equation}
The initial consumption of products in sector $s$ of firm $i$ before the disaster is also defined for convenience:
\begin{equation}
A_{\mbox{tot}i,s}=\Sigma_{j\in s}A_{i,j}.
\end{equation}
The maximum possible production of firm $i$ limited by the inventory of product of sector $s$ on day $t$, $P_{\mbox{pro}i,s}(t)$, is given by
\begin{equation}
P_{\mbox{pro}i,s}(t)=\frac{S_{\mbox{tot}i,s}(t)}{A_{\mbox{tot}i,s}}P_{\mbox{ini}i}.
\end{equation}
Then, we can determine the maximum production of firm $i$ on day $t$, considering its production capacity, $P_{\mbox{cap}i}(t)$, and its production constraints due to the shortage of supplies, $P_{\mbox{pro}i,s}(t)$:
\begin{equation}
P_{\mbox{max}i}(t)=\mbox{Min}\left(P_{\mbox{cap}i}(t), \mbox{Min}_{s}(P_{\mbox{pro}i,s}(t))\right). \label{eq:pconst}
\end{equation}
Therefore, the actual production of firm $i$ on day $t$ is given by
\begin{equation}
P_{\mbox{act}i}(t)=\mbox{Min}\left(P_{\mbox{max}i}(t), D_i(t)\right).
\label{eq:act}
\end{equation}

When demand for a firm is greater than its production capacity, the firm cannot completely satisfy its demand, as denoted by Equation (9). In this case, firms should ration their production to their customers. We propose a rationing policy in which customers and final consumers are prioritised if they have small amount of order to their initial order, instead of being treated equally, as in the previous work~\cite{Hallegatte08}.

Suppose that firm $i$ has customers $j$ and a final consumer. Then, the ratios of the order from customers $j$ and the final consumer after the shock to the one before the shock denoted by $O^{rel}_{j,i}$ and $O^{rel}_{c}$, respectively are determined by the following steps, where $O^{sub}_{j,i}$ and $O^{sub}_{c}$ are temporal variables used to calculate the realised order and are set to be zero initially. 
\begin{enumerate}
\item Obtain the remaining production $r$ of firm $i$
\item Calculate $O^{rel}_{\mbox{min}}=\mbox{Min}(O^{rel}_{j,i}, O^{rel}_{c})$
\item If $r \leq (\sum_{j}{O^{rel}_{\mbox{min}}O_{j,i}}+O^{rel}_{\mbox{min}}C_i)$ then proceed to 8
\item Add $O^{rel}_{\mbox{min}}$ to $O^{sub}_{j,i}$ and $O^{sub}_{c}$ 
\item Subtract $(\sum_{j}{O^{rel}_{\mbox{min}}O_{j,i}}+O^{rel}_{\mbox{min}}C_i)$ from $r$ 
\item Remove the customer or the final consumer that indicated $O^{rel}_{\mbox{min}}$ from the calculation
\item Return to Step 2
\item Calculate $O^{rea}$ that satisfies $r=(\sum_{j}{O^{rea}O_{j,i}}+O^{rea}C_i)$
\item Obtain $O_{j,i}^*=O^{rea}O_{j,i}+O^{sub}_{j,i}O_{j,i}$ and $C_i^*=O^{rea}C_{i}+O^{sub}_{c}C_{i}$,
where the realised order from firm $j$ to supplier $i$ is denoted by $O_{j,i}^* (t)$, and the realised order from a final consumer is $C_i^*$
\item Finalise the calculation
\end{enumerate}


Under this rationing policy, total realised demand for firm $i$, $D_i^* (t)$, is given by
\begin{equation}
D_i^*(t)=\Sigma_jO_{i,j}^*(t)+C_i^*,
\end{equation}
where the realised order from firm $i$ to supplier $j$ is denoted by $O_{i,j}^*(t)$ and that from the final consumers is $C_i^*$. According to firms' production and procurement activities on day $t$, the inventory of firm $j$'s product in firm $i$ on day $t+1$ is updated to
\begin{equation}
S_{i,j}(t+1)=S_{i,j}(t)+O^{*}_{i,j}(t)-A_{i,j}\frac{P_{\mbox{act}i}(t-1)}{P_{\mbox{ini}i}}.
\end{equation}

Several caveats of this model and data should be mentioned. First, we assume that firms cannot find any new supplier when facing a shortage of supplies from their current suppliers. 
Second, for simplicity, our model assumes that inputs from the service sector can be stored as inventory, just like inputs from manufacturing.  
Third, our model ignores changes in the prices of products and wages of labour incorporated in~\cite{Otto2017, Colon2019} and focuses on the dynamics of production because of supply-chain disruptions. 
Fourth, the TSR data report only the location of the headquarters of each firm, not the location of its branches. Because the headquarters of firms are concentrated in Tokyo, production activities in Tokyo are most likely to be overvalued in our analysis. 
Fifth, because of data limitations, we ignore the international supply-chain links in our simulations. 
Finally, this study ignores the impacts of COVID-19 on human and firm behaviours in the post-COVID era. These behavioural changes may influence consumption and production that are assumed to remain the same in this era.

\subsection{Sectoral differences in production capacity after lockdowns}
\label{ch:indmulti}

No data for production capacity (i.e. $P\mbox{cap}$ in the model) during the lockdown of Japan at the firm or sector level are available. Although the Indices of All Industry Activities (IAIA) provides data for {\it post-lockdown production} at the sector level (Section \ref{sec:methodsim}), or $P\mbox{act}$ in our model averaged within a sector, we need information about {\it production capacity}, $P\mbox{cap}$. Therefore, we assume that the rate of reduction in production capacity for each sector is given by the degree of the reduction because of exposure to the virus~\cite{Bonadio2020} multiplied by the share of workers who cannot work at home~\cite{Guan2020} (Section \ref{sec:methodsim}). The rate of reduction because of exposure to the virus is determined by how the workers in the sector have to reduce their activities to avoid contact with others for infection prevention. Because~\cite{Guan2020} define the rate of reduction uniformly worldwide, we modify the rate for some sectors that clearly differ from the practice in Japan. Table~\ref{tbl:mul} shows the rates of reduction for each sector assumed in our simulations.

{\footnotesize
\begin{longtable}{| l | l | l | l | l | l |} 
\caption{Sector-specific rates of reduction in production capacity. Sectors are classified by the JSIC~\cite{MIC2013} at the two-digit level, except for industries 560, 561, and 569 for which we use three-digit codes to reflect the actual circumstances. The sector names are abbreviated. Table~\ref{tbl:abb} lists the sector descriptions and abbreviations.} \label{tbl:mul}\\
\hline
Code & Sector & Reduction rate & Work-at & Exposure & Rationale \\
 & (abbreviated) & & -home rate & level & \\
\hline
\hline
1 & AGR. & 0.433 & 0.134  & 0.5 & Low exposure \\
2 & FRS. & 0.433 & 0.134  & 0.5 & Low exposure \\
3 & FIS. & 0.433 & 0.134  & 0.5 & Low exposure \\
4 & AQA. & 0.433 & 0.134  & 0.5 & Low exposure \\
5 & MIN. & 0.637 & 0.363  & 1 & Ordinary \\
6 & CNS.GEN. & 0.758 & 0.242  & 1 & Ordinary \\
7 & CNS.SPC. & 0.758 & 0.242  & 1 & Ordinary \\
8 & EQP. & 0.758 & 0.242  & 1 & Ordinary \\
9 & MAN.FOD. & 0.76 & 0.240  & 1 & Ordinary \\
10 & MAN.BEV. & 0.76 & 0.240  & 1 & Ordinary \\
11 & MAN.TEX & 0.668 & 0.332  & 1 & Ordinary \\
12 & MAN.LUM. & 0.768 & 0.232  & 1 & Ordinary \\
13 & MAN.FUR. & 0.768 & 0.232  & 1 & Ordinary \\
14 & MAN.PUL. & 0.676 & 0.324  & 1 & Ordinary \\
15 & PRT. & 0.676 & 0.324  & 1 & Ordinary \\
16 & MAN.CHM. & 0.529 & 0.471  & 1 & Ordinary \\
17 & MAN.PET. & 0.651 & 0.349  & 1 & Ordinary \\
18 & MAN.PLA. & 0.704 & 0.296  & 1 & Ordinary \\
19 & MAN.RUB. & 0.704 & 0.296  & 1 & Ordinary \\
20 & MAN.LET. & 0.668 & 0.332  & 1 & Ordinary \\
21 & MAN.CER. & 0.709 & 0.291  & 1 & Ordinary \\
22 & MAN.IRN. & 0.732 & 0.268  & 1 & Ordinary \\
23 & MAN.NFM. & 0.732 & 0.268  & 1 & Ordinary \\
24 & MAN.FBM. & 0.695 & 0.305  & 1 & Ordinary \\
25 & MAN.GNM. & 0.604 & 0.396  & 1 & Ordinary \\
26 & MAN.PRM. & 0.604 & 0.396  & 1 & Ordinary \\
27 & MAN.BSM. & 0.604 & 0.396  & 1 & Ordinary \\
28 & EPT. & 0.333 & 0.667  & 1 & Ordinary \\
29 & MAN.ELM. & 0.58 & 0.420  & 1 & Ordinary \\
30 & MAN.INF. & 0.333 & 0.667  & 1 & Ordinary \\
31 & MAN.TRN. & 0.504 & 0.496  & 1 & Ordinary \\
32 & MAN.MSC. & 0.705 & 0.295  & 1 & Ordinary \\
33 & ELE. & 0.0623 & 0.377  & 0.1 & Lifeline \\
34 & GAS. & 0.0623 & 0.377  & 0.1 & Lifeline \\
35 & HET. & 0.0623 & 0.377  & 0.1 & Lifeline \\
36 & WTR. & 0.0623 & 0.377  & 0.1 & Lifeline \\
37 & COM. & 0.0401 & 0.599  & 0.1 & Lifeline \\
38 & BRD. & 0.0192 & 0.808  & 0.1 & Lifeline \\
39 & INF.SVC. & 0.097 & 0.903  & 1 & Ordinary \\
40 & INT. & 0.0401 & 0.599  & 0.1 & Lifeline \\
41 & INF.DST. & 0.192 & 0.808  & 1 & Ordinary \\
42 & RLW.TRP. & 0.0701 & 0.299  & 0.1 & Lifeline \\
43 & PAS.TRP. & 0.0701 & 0.299  & 0.1 & Lifeline \\
44 & FRE.TRP. & 0.0701 & 0.299  & 0.1 & Lifeline \\
45 & WTR.TRP. & 0.0701 & 0.299  & 0.1 & Lifeline \\
46 & AIR.TRP. & 0.0701 & 0.299  & 0.1 & Lifeline \\
47 & WRH. & 0.0701 & 0.299  & 0.1 & Lifeline \\
48 & SVC.TRP. & 0.0701 & 0.299  & 0.1 & Lifeline \\
49 & PST.SVC. & 0.0701 & 0.299  & 0.1 & Lifeline \\
50 & WHL.GEN. & 0.525 & 0.475  & 1 & Ordinary \\
51 & WHL.TEX. & 0.525 & 0.475  & 1 & Ordinary \\
52 & WHL.FOD. & 0.525 & 0.475  & 1 & Ordinary \\
53 & WHL.MAT. & 0.525 & 0.475  & 1 & Ordinary \\
54 & WHL.MCN. & 0.525 & 0.475  & 1 & Ordinary \\
55 & WHL.MSC. & 0.525 & 0.475  & 1 & Ordinary \\
560 & RTL.ADM. & 0.525 & 0.475  & 1 & Ordinary \\
561 & RTL.DPT. & 0.525 & 0.475  & 1 & Closed \\
569 & RTL.GNM. & 0.0525 & 0.475  & 0.1 & Lifeline \\
57 & RTL.GEN. & 0.525 & 0.475  & 1 & Ordinary \\
58 & RTL.FOD. & 0.525 & 0.475  & 1 & Ordinary \\
59 & RTL.MCN. & 0.525 & 0.475  & 1 & Ordinary \\
60 & RTL.MSC. & 0.525 & 0.475  & 1 & Ordinary \\
61 & RTL.NST. & 0.525 & 0.475  & 1 & Ordinary \\
62 & FIN.BNK. & 0.214 & 0.786  & 1 & Ordinary \\
63 & FIN.ORG. & 0.214 & 0.786  & 1 & Ordinary \\
64 & FIN.LON. & 0.214 & 0.786  & 1 & Ordinary \\
65 & FIN.TRN. & 0.214 & 0.786  & 1 & Ordinary \\
66 & FIN.AUX. & 0.214 & 0.786  & 1 & Ordinary \\
67 & INS. & 0.214 & 0.786  & 1 & Ordinary \\
68 & RST.AGN. & 0.423 & 0.577  & 1 & Ordinary \\
69 & RTS.LES. & 0.423 & 0.577  & 1 & Ordinary \\
70 & RNT. & 0.362 & 0.638  & 1 & Ordinary \\
71 & SCI. & 0.172 & 0.828  & 1 & Ordinary \\
72 & SVC.PRF. & 0.362 & 0.638  & 1 & Ordinary \\
73 & ADV. & 0.362 & 0.638  & 1 & Ordinary \\
74 & SVC.TEC. & 0.362 & 0.638  & 1 & Ordinary \\
75 & ACM. & 0.889 & 0.111  & 1 & Closed \\
76 & EAT. & 0.889 & 0.111  & 1 & Ordinary \\
77 & DEL. & 0.0521 & 0.479  & 0.1 & Lifeline \\
78 & LND. & 0.521 & 0.479  & 1 & Ordinary \\
79 & SVC.PSN. & 0.521 & 0.479  & 1 & Ordinary \\
80 & SVC.AMS. & 0.521 & 0.479  & 1 & Closed \\
81 & SCH. & 0.086 & 0.828  & 0.5 & Low exposure \\
82 & EDC. & 0.086 & 0.828  & 0.5 & Low exposure \\
83 & MED. & 0.0753 & 0.247  & 0.1 & Lifeline \\
84 & HLT. & 0 & 0.247  & 0 & Sustantial \\
85 & WEL. & 0 & 0.247  & 0 & Sustantial \\
86 & PST.OFC. & 0.0362 & 0.638  & 0.1 & Lifeline \\
87 & CAS. & 0.181 & 0.638  & 0.5 & Low exposure \\
88 & WAS. & 0.181 & 0.638  & 0.5 & Low exposure \\
89 & SVC.AUT. & 0.181 & 0.638  & 0.5 & Low exposure \\
90 & SVC.MCN. & 0.181 & 0.638  & 0.5 & Low exposure \\
91 & SVC.EMP. & 0.181 & 0.638  & 0.5 & Low exposure \\
92 & SVC.BUS. & 0.181 & 0.638  & 0.5 & Low exposure \\
93 & PLT. & 0.181 & 0.638  & 0.5 & Low exposure \\
94 & REL. & 0.181 & 0.638  & 0.5 & Low exposure \\
95 & SVC.MSC. & 0.181 & 0.638  & 0.5 & Low exposure \\
96 & GOV.INT. & 0.0515 & 0.485  & 0.1 & Lifeline \\
97 & NA & 0.0515 & 0.485  & 0.1 & Lifeline \\
98 & GOV.LOC. & 0.0515 & 0.485  & 0.1 & Lifeline \\
99 & NEC & 0.362 & 0.638  & 1 & Ordinary \\

\hline
\end{longtable}
}

\begin{landscape}
{\footnotesize
\begin{longtable}{| l | l | l |} 
\caption{Sector classifications and abbreviations.} \label{tbl:abb}\\
\hline
Code & Description & Abbreviation \\
\hline
\hline
01 & AGRICULTURE & AGR. \\
02 & FORESTRY & FRS. \\
03 & FISHERIES, EXCEPT AQUACULTURE & FIS. \\
04 & AQUACULTURE & AQA. \\
05 & MINING AND QUARRYING OF STONE AND GRAVEL & MIN. \\
06 & CONSTRUCTION WORK, GENERAL INCLUDING PUBLIC AND PRIVATE CONSTRUCTION WORK & CNS.GEN. \\
07 & CONSTRUCTION WORK BY SPECIALIST CONTRACTOR, EXCEPT EQUIPMENT INSTALLATION WORK & CNS.SPC. \\
08 & EQUIPMENT INSTALLATION WORK & EQP. \\
09 & MANUFACTURE OF FOOD & MAN.FOD. \\
10 & MANUFACTURE OF BEVERAGES, TOBACCO AND FEED & MAN.BEV. \\
11 & MANUFACTURE OF TEXTILE PRODUCTS & MAN.TEX \\
12 & MANUFACTURE OF LUMBER AND WOOD PRODUCTS, EXCEPT FURNITURE & MAN.LUM. \\
13 & MANUFACTURE OF FURNITURE AND FIXTURES & MAN.FUR. \\
14 & MANUFACTURE OF PULP, PAPER AND PAPER PRODUCTS & MAN.PUL. \\
15 & PRINTING AND ALLIED INDUSTRIES & PRT. \\
16 & MANUFACTURE OF CHEMICAL AND ALLIED PRODUCTS & MAN.CHM. \\
17 & MANUFACTURE OF PETROLEUM AND COAL PRODUCTS & MAN.PET. \\
18 & MANUFACTURE OF PLASTIC PRODUCTS, EXCEPT OTHERWISE CLASSIFIED & MAN.PLA. \\
19 & MANUFACTURE OF RUBBER PRODUCTS & MAN.RUB. \\
20 & MANUFACTURE OF LEATHER TANNING, LEATHER PRODUCTS AND FUR SKINS & MAN.LET. \\
21 & MANUFACTURE OF CERAMIC, STONE AND CLAY PRODUCTS & MAN.CER. \\
22 & MANUFACTURE OF IRON AND STEEL & MAN.IRN. \\
23 & MANUFACTURE OF NON-FERROUS METALS AND PRODUCTS & MAN.NFM. \\
24 & MANUFACTURE OF FABRICATED METAL PRODUCTS & MAN.FBM. \\
25 & MANUFACTURE OF GENERAL-PURPOSE MACHINERY & MAN.GNM. \\
26 & MANUFACTURE OF PRODUCTION MACHINERY & MAN.PRM. \\
27 & MANUFACTURE OF BUSINESS ORIENTED MACHINERY & MAN.BSM. \\
28 & ELECTRONIC PARTS, DEVICES AND ELECTRONIC CIRCUITS & EPT. \\
29 & MANUFACTURE OF ELECTRICAL MACHINERY, EQUIPMENT AND SUPPLIES & MAN.ELM. \\
30 & MANUFACTURE OF INFORMATION AND COMMUNICATION ELECTRONICS EQUIPMENT & MAN.INF. \\
31 & MANUFACTURE OF TRANSPORTATION EQUIPMENT & MAN.TRN. \\
32 & MISCELLANEOUS MANUFACTURING INDUSTRIES & MAN.MSC. \\
33 & PRODUCTION, TRANSMISSION AND DISTRIBUTION OF ELECTRICITY & ELE. \\
34 & PRODUCTION AND DISTRIBUTION OF GAS & GAS. \\
35 & HEAT SUPPLY & HET. \\
36 & COLLECTION, PURIFICATION AND DISTRIBUTION OF WATER, AND SEWAGE COLLECTION, PROCESSING & WTR. \\
37 & COMMUNICATIONS & COM. \\
38 & BROADCASTING & BRD. \\
39 & INFORMATION SERVICES & INF.SVC. \\
40 & SERVICES INCIDENTAL TO INTERNET & INT. \\
41 & VIDEO PICTURE INFORMATION, SOUND INFORMATION, CHARACTER INFORMATION PRODUCTION AND DISTRIBUTION & INF.DST. \\
42 & RAILWAY TRANSPORT & RLW.TRP. \\
43 & ROAD PASSENGER TRANSPORT & PAS.TRP. \\
44 & ROAD FREIGHT TRANSPORT & FRE.TRP. \\
45 & WATER TRANSPORT & WTR.TRP. \\
46 & AIR TRANSPORT & AIR.TRP. \\
47 & WAREHOUSING & WRH. \\
48 & SERVICES INCIDENTAL TO TRANSPORT & SVC.TRP. \\
49 & POSTAL SERVICES, INCLUDING MAIL DELIVERY & PST.SVC. \\
50 & WHOLESALE TRADE, GENERAL MERCHANDISE & WHL.GEN. \\
51 & WHOLESALE TRADE (TEXTILE AND APPAREL) & WHL.TEX. \\
52 & WHOLESALE TRADE (FOOD AND BEVERAGES) & WHL.FOD. \\
53 & WHOLESALE TRADE (BUILDING MATERIALS, MINERALS AND METALS, ETC) & WHL.MAT. \\
54 & WHOLESALE TRADE (MACHINERY AND EQUIPMENT) & WHL.MCN. \\
55 & MISCELLANEOUS WHOLESALE TRADE & WHL.MSC. \\
560 & ESTABLISHMENTS ENGAGED IN ADMINISTRATIVE OR ANCILLARY ECONOMIC ACTIVITIES & RTL.ADM. \\
561 & DEPARTMENT STORES AND GENERAL MERCHANDISE SUPERMARKET & RTL.DPT. \\
569 & MISCELLANEOUS RETAIL TRADE, GENERAL MERCHANDISE  & RTL.GNM. \\
57 & RETAIL TRADE, GENERAL MERCHANDISE & RTL.GEN. \\
58 & RETAIL TRADE (FOOD AND BEVERAGE) & RTL.FOD. \\
59 & RETAIL TRADE (MACHINERY AND EQUIPMENT) & RTL.MCN. \\
60 & MISCELLANEOUS RETAIL TRADE & RTL.MSC. \\
61 & NONSTORE RETAILERS & RTL.NST. \\
62 & BANKING & FIN.BNK. \\
63 & FINANCIAL INSTITUTIONS FOR COOPERATIVE ORGANIZATIONS & FIN.ORG. \\
64 & NON-DEPOSIT MONEY CORPORATIONS, INCLUDING LENDING AND CREDIT CARD BUSINESS & FIN.LON. \\
65 & FINANCIAL PRODUCTS TRANSACTION DEALERS AND FUTURES COMMODITY TRANSACTION DEALERS & FIN.TRN. \\
66 & FINANCIAL AUXILIARIES & FIN.AUX. \\
67 & INSURANCE INSTITUTIONS, INCLUDING INSURANCE AGENTS, BROKERS AND SERVICES & INS. \\
68 & REAL ESTATE AGENCIES & RST.AGN. \\
69 & REAL ESTATE LESSORS AND MANAGERS & RTS.LES. \\
70 & GOODS RENTAL AND LEASING & RNT. \\
71 & SCIENTIFIC AND DEVELOPMENT RESEARCH INSTITUTES & SCI. \\
72 & PROFESSIONAL SERVICES, N.E.C. & SVC.PRF. \\
73 & ADVERTISING & ADV. \\
74 & TECHNICAL SERVICES, N.E.C. & SVC.TEC. \\
75 & ACCOMMODATION & ACM. \\
76 & EATING AND DRINKING PLACES & EAT. \\
77 & FOOD TAKE OUT AND DELIVERY SERVICES & DEL. \\
78 & LAUNDRY, BEAUTY AND BATH SERVICES & LND. \\
79 & MISCELLANEOUS LIVING-RELATED AND PERSONAL SERVICES & SVC.PSN. \\
80 & SERVICES FOR AMUSEMENT AND RECREATION & SVC.AMS. \\
81 & SCHOOL EDUCATION & SCH. \\
82 & MISCELLANEOUS EDUCATION, LEARNING SUPPORT & EDC. \\
83 & MEDICAL AND OTHER HEALTH SERVICE & MED. \\
84 & PUBLIC HEALTH AND HYGIENE & HLT. \\
85 & SOCIAL INSURANCE, SOCIAL WELFARE AND CARE SERVICES & WEL. \\
86 & POSTAL OFFICE & PST.OFC. \\
87 & COOPERATIVE ASSOCIATIONS, N.E.C. & CAS. \\
88 & WASTE DISPOSAL BUSINESS & WAS. \\
89 & AUTOMOBILE MAINTENANCE SERVICES & SVC.AUT. \\
90 & MACHINE, ETC. REPAIR SERVICES, EXCEPT OTHERWISE CLASSIFIED & SVC.MCN. \\
91 & EMPLOYMENT AND WORKER DISPATCHING SERVICES & SVC.EMP. \\
92 & MISCELLANEOUS BUSINESS SERVICES & SVC.BUS. \\
93 & POLITICAL, BUSINESS AND CULTURAL ORGANIZATIONS & PLT. \\
94 & RELIGION & REL. \\
95 & MISCELLANEOUS SERVICES & SVC.MSC. \\
96 & FOREIGN GOVERNMENTS AND INTERNATIONAL AGENCIES IN JAPAN & GOV.INT. \\
97 & NATIONAL GOVERNMENT SERVICES & GOV.NAT. \\
98 & LOCAL GOVERNMENT SERVICES & GOV.LOC. \\
99 & INDUSTRIES UNABLE TO CLASSIFY & NEC \\
\hline
\end{longtable}
}
\end{landscape}

\subsection{Helmholtz-Hodge decomposition}
\label{ch:hhd}


The Helmholtz-Hodge decomposition (HHD) decomposes a flow from a node to another in a network into a potential flow component and a loop flow component. A potential flow component is determined by the upstream/downstream location of the node in a network~\cite{jiang2011hodge}, whereas a loop flow component is given by a constraint such that the summation of the incoming and outgoing loop flows of all the nodes equals zero. This method has been used to find the structure of potential and loop flows in complex networks. See, for example, \cite{kichikawa2019community,scirep2020,mackay2020directed,Fujiwara2020}.

Suppose we have a flow of a matrix denoted by $B_{ij}$ such that a flow from node $i$ to node $j$ is represented by $B_{ij}$. For simplicity, we assume $\forall i,j~B_{ij}\geq 0 $. $A_{ij}$ is a binary adjacency matrix generated from $B_{ij}$:
\begin{eqnarray}
  \label{eq:def_Aij}
  A_{ij} = & 1 & \mbox{if $B_{ij}>0$}, \nonumber \\
    & 0 & \mbox{otherwise}.
\end{eqnarray}
We define a `net flow' $F_{ij}$ by
\begin{equation}
  \label{eq:def_Fij}
  F_{ij}=B_{ij}-B_{ji},
\end{equation}
and a `net weight' $w_{ij}$ by
\begin{equation}
  \label{eq:def_wij}
  w_{ij}=A_{ij}+A_{ji}.
\end{equation}
Note that $w_{ij}$ is symmetric, $w_{ij}= w_{ji}$, and non-negative, $w_{ij}\geq 0$, for any pair of $i$ and $j$. 

Then, the HHD is given by
\begin{equation}
  \label{eq:def_hodge}
  F_{ij}=F^{(c)}_{ij}+F^{(p)}_{ij},
\end{equation}
where the loop flow $F^{(c)}_{ij}$ satisfies
\begin{equation}
  \label{eq:def_Fcirc}
  \sum_j  F^{(c)}_{ij}=0,
\end{equation}
meaning that loop flows are divergence-free.
The potential flow, $F^{(p)}_{ij}$, can be expressed as
\begin{equation}
  \label{eq:def_Fgrad}
  F^{(p)}_{ij}=w_{ij}(\phi_i - \phi_j),
\end{equation}
where $\phi_i$ is the Helmholtz-Hodge potential of node $i$ that identifies its upstream/downstream position in the network. More precisely, $\phi_i$ is larger when node $i$ is located in a more upstream position in the network and vice versa.
Equation~(\ref{eq:def_Fgrad}) indicates that the potential flow $F^{(p)}_{ij}$ is the difference in the HH potential between two nodes when the two are linked and zero when they are not linked.  We further assume 
\begin{equation}
    \label{eq:pot_sum}
    \sum_i \phi_i = 0
\end{equation}
for normalisation purposes. Then, equations (\ref{eq:def_hodge})--(\ref{eq:pot_sum}) can be uniquely solved for $F_{ij}^{(c)}$, $F_{ij}^{(p)}$, and $\phi_i$ for all $i$ and $j$ in the whole network.

\begin{figure}[htb]
\centering
\includegraphics[width=.7\linewidth]{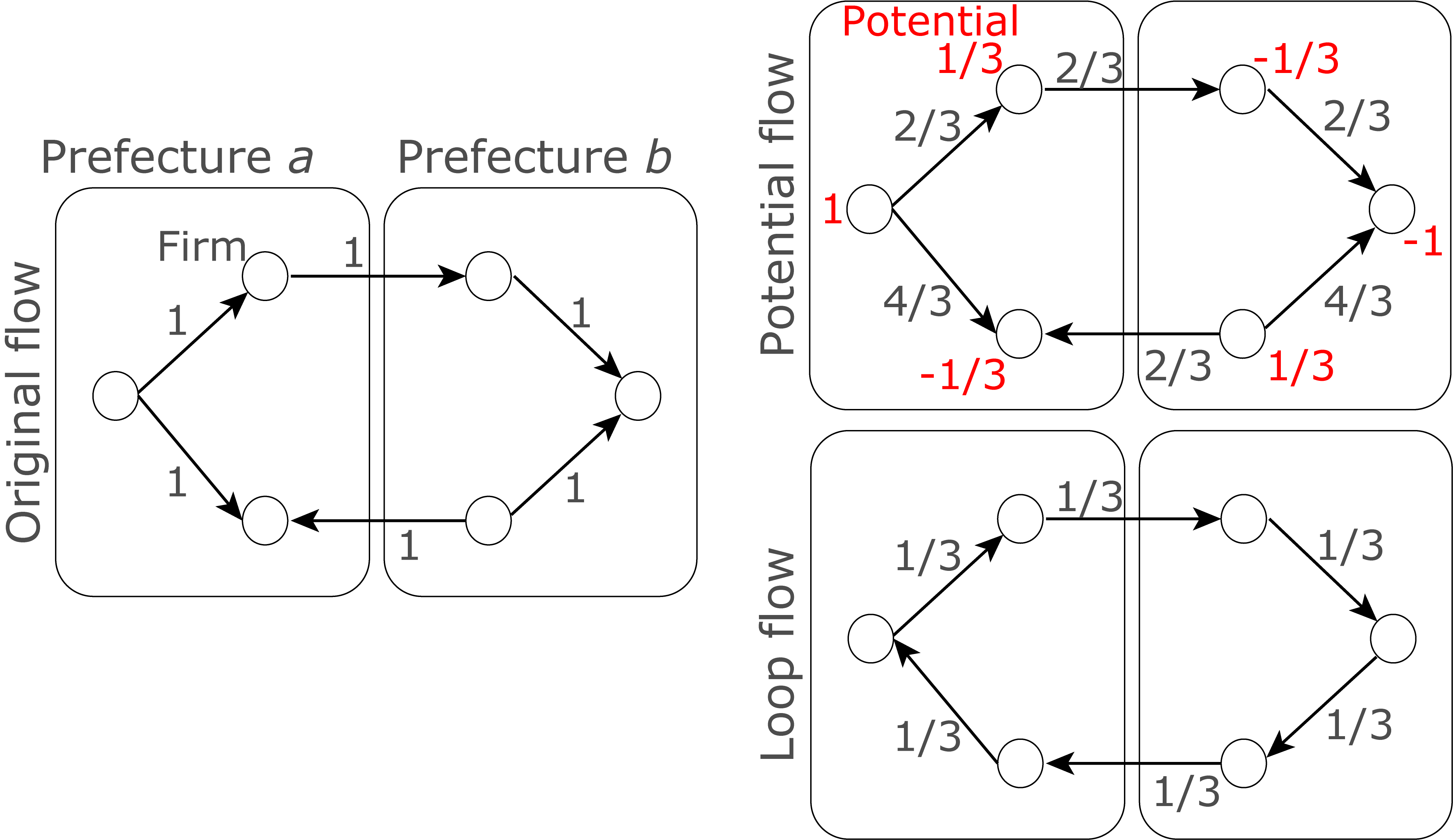}
\caption{An example of the HHD and loop and potential flow measures of prefectures. The left panel shows the supply chains of the six firms in the two prefectures. The right top and bottom panels present the potential flows and loop flows, respectively obtained from the HHD.
}
\label{fig:hhdponchi}
\end{figure}

Figure~\ref{fig:hhdponchi} shows a simple example to explain the intuition behind the potential and loop flows, potential obtained from the HHD, and potential and loop flow measures between two prefectures (i.e. $Pot_{ab}$, $Pot_{ba}$, and $Loop_{ab}$ defined in Section~\ref{ch:lifttwo}). 
The left panel shows a supply chain with six firms in prefectures $a$ and $b$. The right top and bottom panels indicate the potential flows and loop flows, respectively decomposed by the HHD. The numbers in red in the right top panel represent the HH potential, or the upstreamness in supply chains, for each firm.  
Although there is no `loop' in a standard sense among the firms in this example, the HHD identifies loop flows in the sense that the nodes in the loop are affected by each other. Hence, shocks circulate in the loop and work differently from those in the non-loop potential flows. 

Specifically, $Pot_{ab}$ is the sum of the total potential flows from the firms in prefecture $a$ to those in prefecture $b$ (there is only a potential flow from prefectures $a$ to $b$ in this example) divided by the total flows of firms in prefecture $a$. Therefore, $Pot_{ab}=(2/3)/4=1/6$. $Pot_{ba}$ is the opposite direction and $Pot_{ba}=1/6$. Because $Loop_{ab}$ is the sum of the total loop flows between the firms in prefectures $a$ and $b$ (there are two loop flows between $a$ and $b$ in this example), $Loop_{ab}=(2/3)/4=1/6$ and, similarly, $Loop_{ba}=1/6$.

\begin{figure}[htb]
\centering
\includegraphics[width=.7\linewidth]{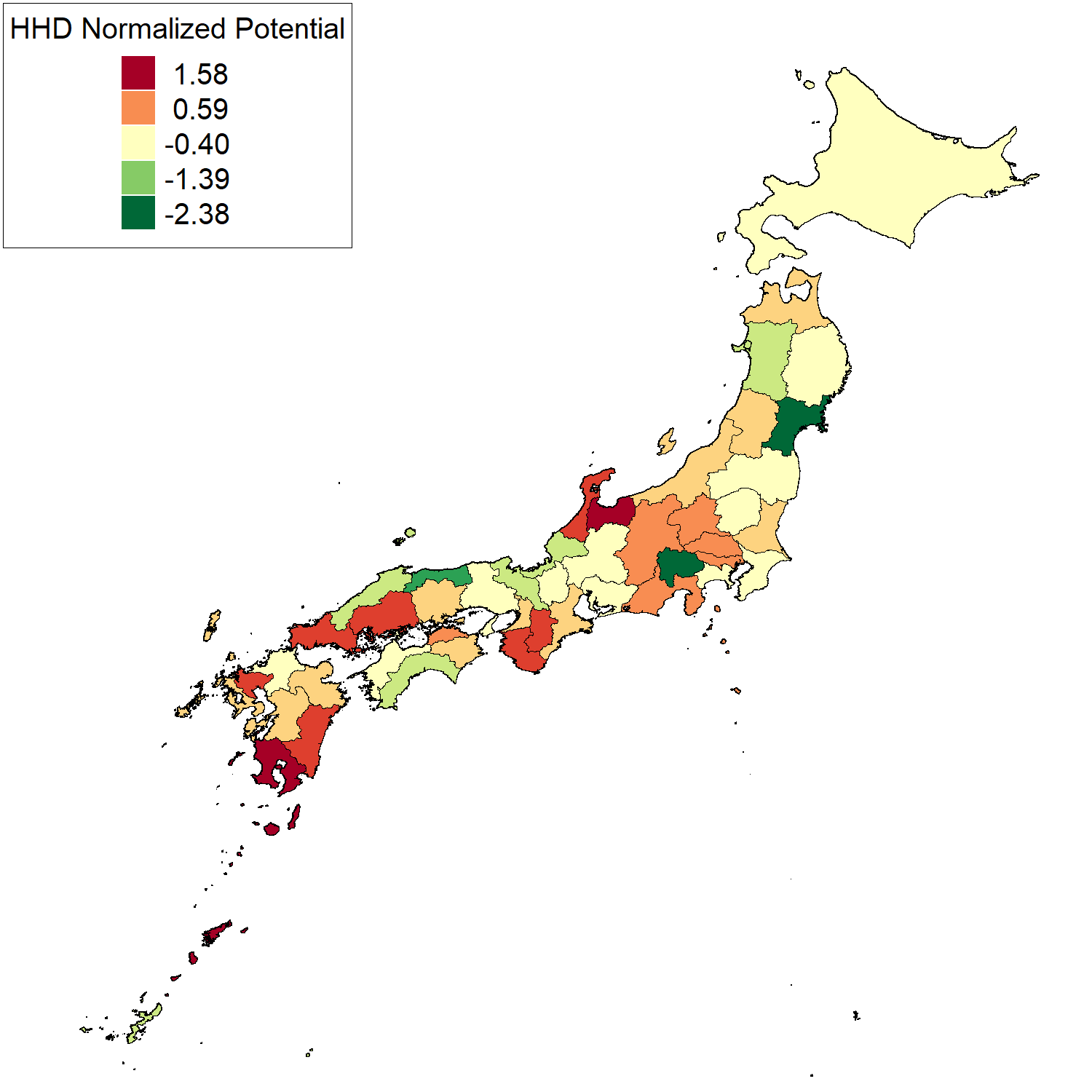}
\caption{Choropleth map of the potential calculated by the Helmholtz-Hodge decomposition. The average HH potential over all the firms in each prefecture is presented.
}
\label{fig:hhdmap}
\end{figure}

Figure~\ref{fig:hhdmap} shows the average of the HH potential $\phi_i$ of the firms in the supply-chain network, which is normalised so that its overall average is zero, for each prefecture. This figure illustrates the large variation in the upstreamness of the firms at the prefecture level. 

\subsection{Substitutability for two regions}
\label{ch:subs}

\begin{figure}[htb]
\centering
\includegraphics[width=.7\linewidth]{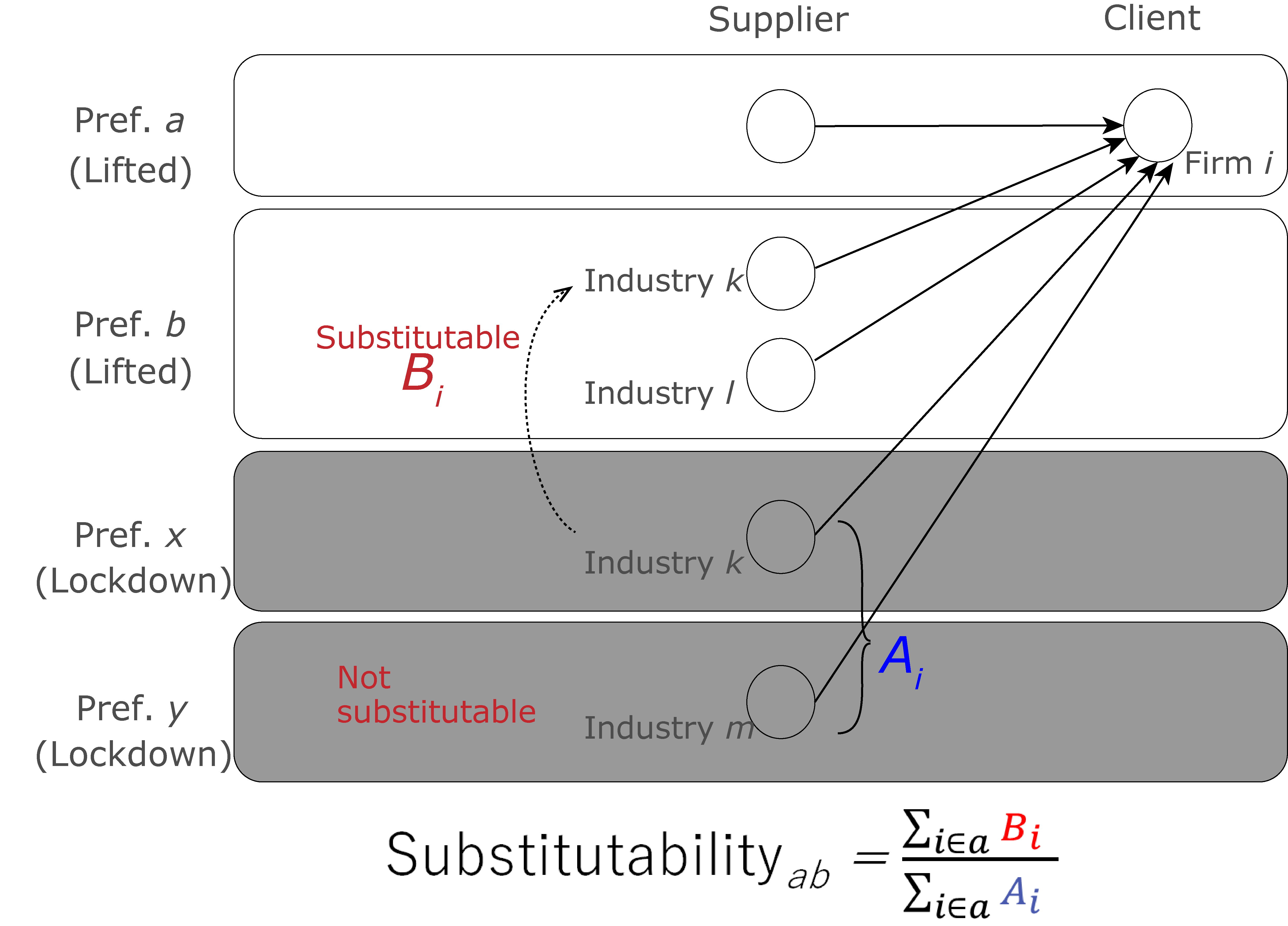}
\caption{An example of the substitutability measure for two regions. The bottom shows the equation. $A_i$ is the total number of suppliers outside prefectures $a$ and $b$. The bottom two suppliers are applicable. A supplier in prefecture $b$ belongs to the same industry as the upper firm of the outside suppliers, whereas the lower firm of the outside suppliers is not substitutable. Hence, $A_i=2$ and $B_i=1$.
}
\label{fig:subs}
\end{figure}

Since the definition of the substitutability measure for two regions is not as simple as
the definition for one region,
we provide a further explanation.
Figure~\ref{fig:subs} is an example for the suppliers of a firm in prefecture $a$.
The substitutability of prefecture $a$ by prefecture $b$
is a fraction.
The denominator is the total number of suppliers
that delivers goods to the firms in prefecture $a$
except suppliers in prefecture $a$ or $b$. (Here, we call this $A_i$ in the figure.)
Hereafter, a supplier means a supplier of a firm in prefecture $a$.
The numerator is the total number of
substitutable suppliers in $A_i$.
A supplier in $A_i$ is substitutable
if a supplier in prefecture $b$ belongs to the same industry as the focal supplier.

\section{Results}

\subsection{Simulation of the effect of the actual lockdown}
\label{ch:Asimu}

A video is available for the temporal and geographical visualisation of the lockdown simulation at 
\url{https://youtu.be/q029a_e1akU}.
The map if the video indicates the rate of reduction in firm production averaged within each municipality. The red areas indicate that the production in the area is less than or equal to 20\% of firms' capacity on average, whereas the light red and orange areas show firms with a more moderate decline in production.
The inset in the video indicates Figure~\ref{fig:actual} and the number of days from the first lockdown.
The visualisation clearly shows the areas that are not locked down are also affected by lockdowns of other areas.
For example, from day 0 to day 8, only seven prefectures are locked down but most of the areas in Japan are affected
(see Section~\ref{sec:soe} and Figure~\ref{fig:mapsoe}).
This reduction of the production happens because
the demand reduction propagates to the suppliers without any buffer. On the other hand, the supply reduction can be mitigated because each client holds inventories for the intermediate goods.

\subsection{Estimation of daily GDP from IAIA}
\label{ch:IAIA}

The IAIA indicates the changes in production in all industries in Japan compared with that in the previous month and in the same month in the previous year, based on firm surveys~\cite{METIIIP2020}. We assume that daily production on 7 April (day 0) is the same as that in March and thus can be calculated from the IAIA in March. Then, we estimate the daily GDP in April (or May) by (yearly GDP)/365$\times$(IAIA in April (May))/(IAIA in March) and illustrate it by the left (right) red line in Figure~\ref{fig:actual}.

\subsection{Interconnected effect of the different strictness of regional lockdowns}
\label{ch:strength}
\setcounter{figure}{0} 
\setcounter{table}{0} 

\begin{figure}[htb]
\centering
\includegraphics[width=0.8\linewidth]{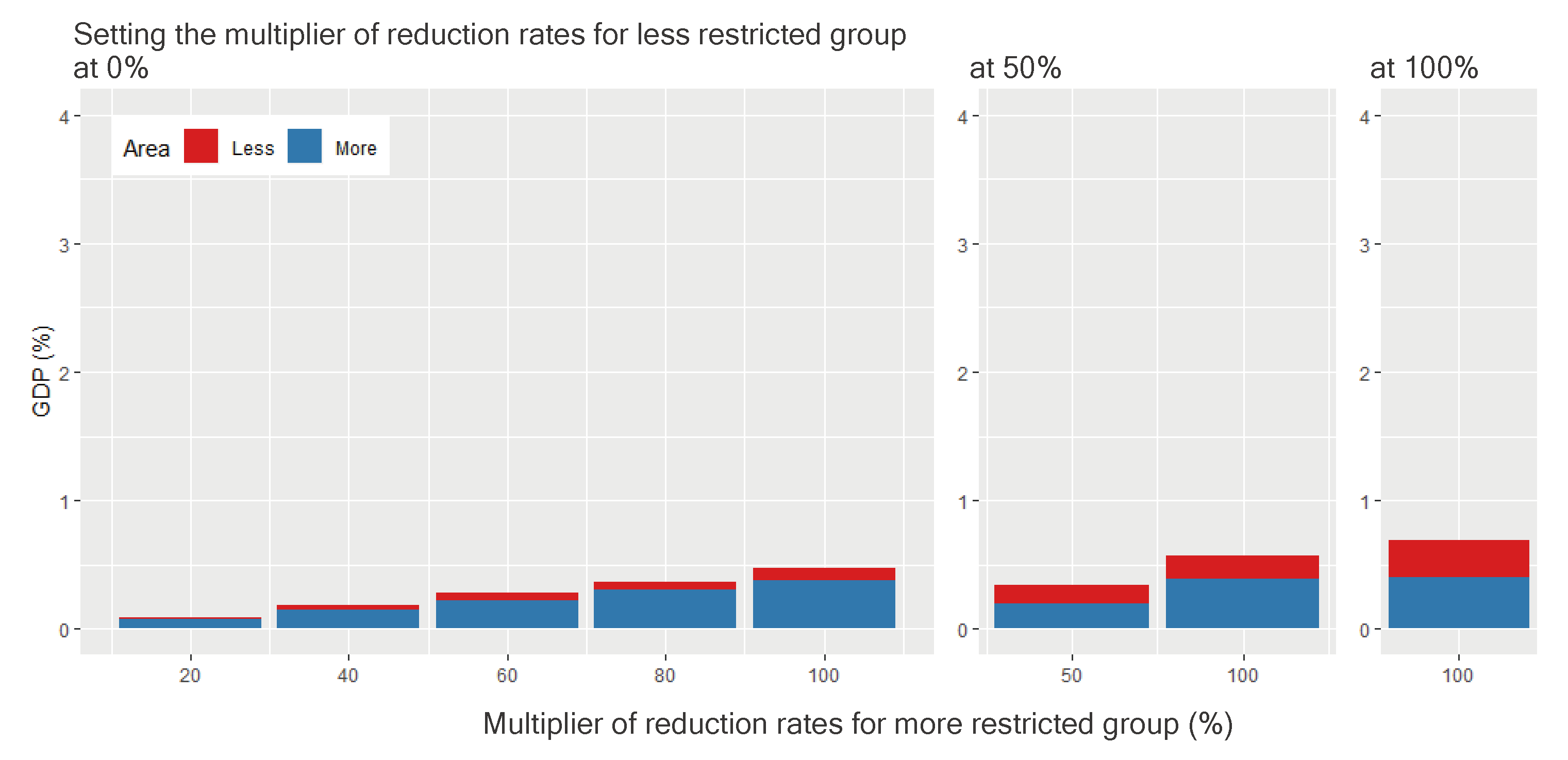}
\caption{Loss in value added as a percentage of total GDP assuming different restriction levels for a lockdown of 14 days between the more and less restricted groups. A restriction level is defined by a multiplier for the sector-specific benchmark rates of reduction in production capacity. The red and blue parts of each bar show the loss of value added in the less and more restricted groups, respectively as a percentage of GDP.}
\label{fig:stack14}
\end{figure}

\begin{figure}[htb]
\centering
\includegraphics[width=0.8\linewidth]{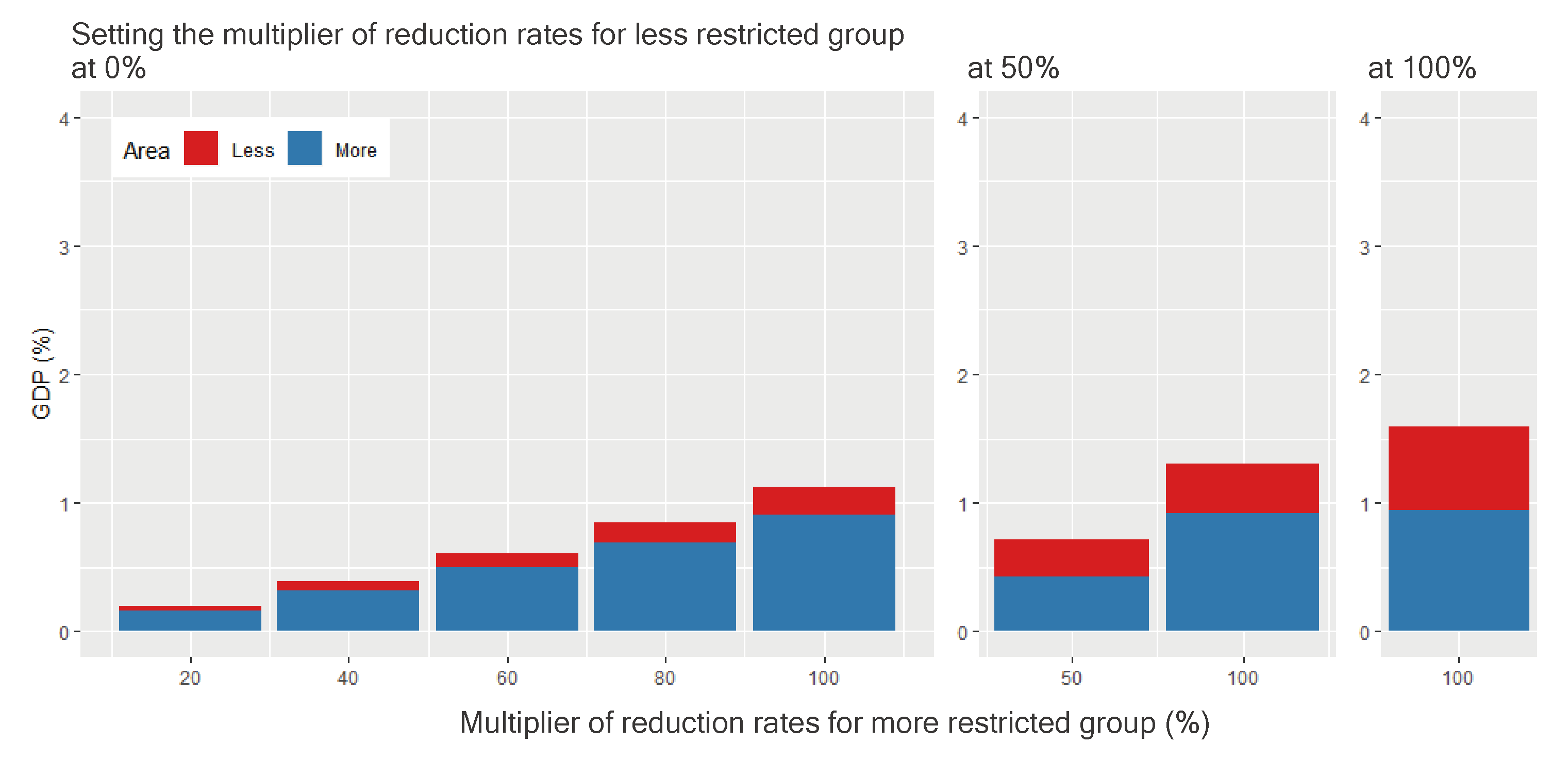}
\caption{Loss in value added as a percentage of total GDP assuming different restriction levels for a lockdown of 30 days between the more and less restricted groups. A restriction level is defined by a multiplier for the sector-specific benchmark rates of reduction in production capacity. The red and blue parts of each bar show the loss of value added in the less and more restricted groups, respectively as a percentage of GDP.}
\label{fig:stack30}
\end{figure}

In Section \ref{ch:sensitivity}, we showed that the different lockdown strictness between the more and less restricted groups affects the economic losses of the two groups, particularly assuming that the lockdown continues for 60 days. We also experiment with different lockdown durations (i.e. 14 and 30 days) and show the results in Figures \ref{fig:stack14} and \ref{fig:stack30}. The main result that the strictness of the lockdown in the more restricted group that includes the major industrial clusters substantially affects the economic loss of the other group by propagation through supply chains still holds. 

\subsection{Effect of lifting the lockdown in one region}
\label{ch:exemption}

\begin{figure}[htb]
\centering
\includegraphics[width=0.8\linewidth]{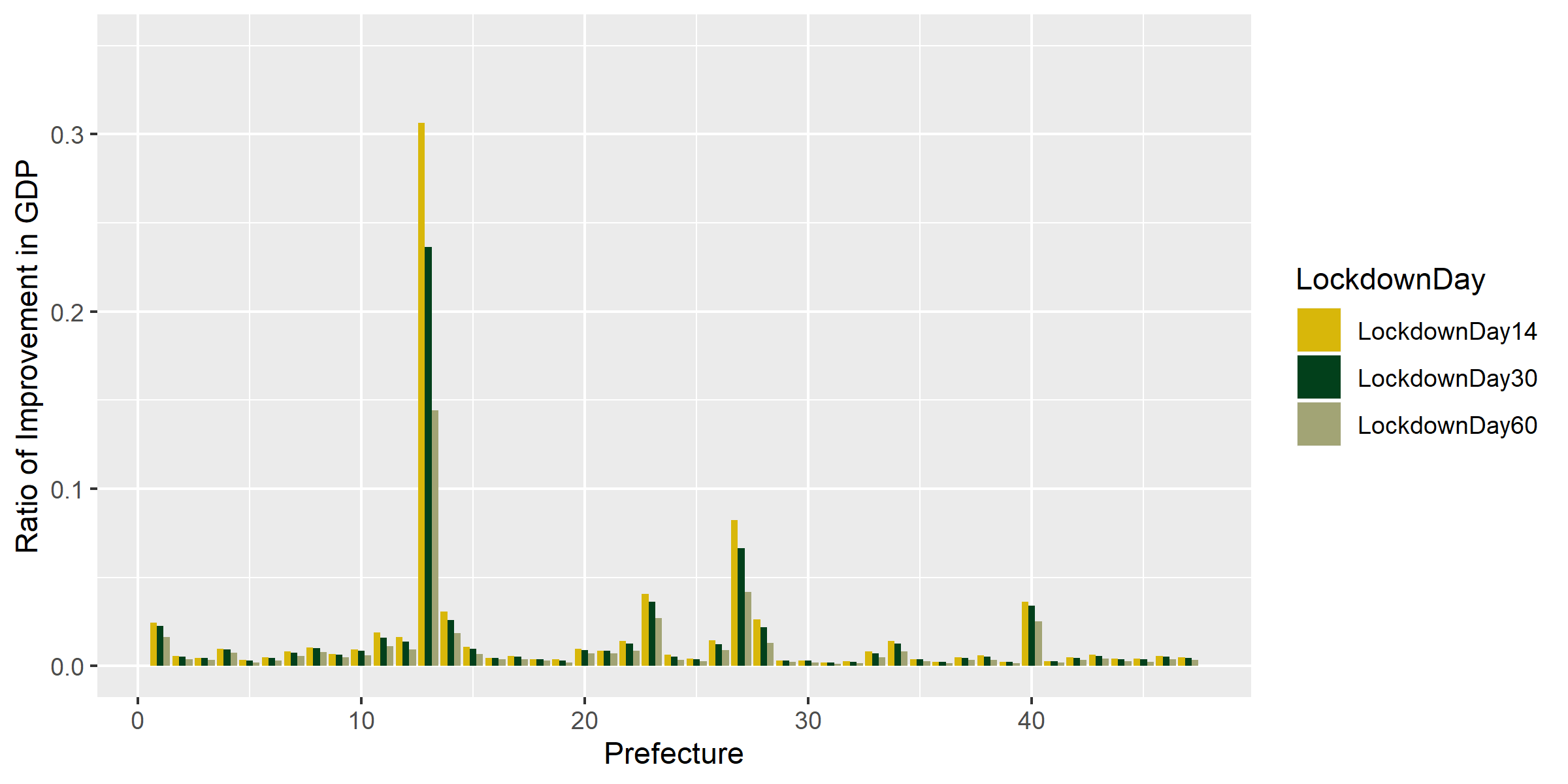}
\caption{
Ratio of the improvement in GDP by lifting the lockdown in each prefecture. The improvement is defined as the ratio of the increase in national GDP by each prefecture lifting its lockdown to the decrease in GDP by all prefectures' lockdowns. The horizontal axis indicates the JIS codes of the prefectures. The yellow, dark green, and light green bars show the ratio of the improvement when lockdowns persist for 14, 30, and 60 days, respectively. 
}
\label{fig:onoffoverall}
\end{figure}

Section \ref{ch:lift} presents the effect of lifting the lockdown in a prefecture on its production, assuming that all the other prefectures are still locked down. Figure~\ref{fig:onoffoverall} shows the ratio of the increase in national GDP from each prefecture lifting its lockdown to the decrease in GDP by all prefectures' lockdowns. The prefectures are horizontally aligned in order of JIS cods. The top three prefectures in terms of the recovery rate are Tokyo, Osaka, and Fukuoka.

\begin{figure}[htb]
\centering
\includegraphics[width=0.8\linewidth]{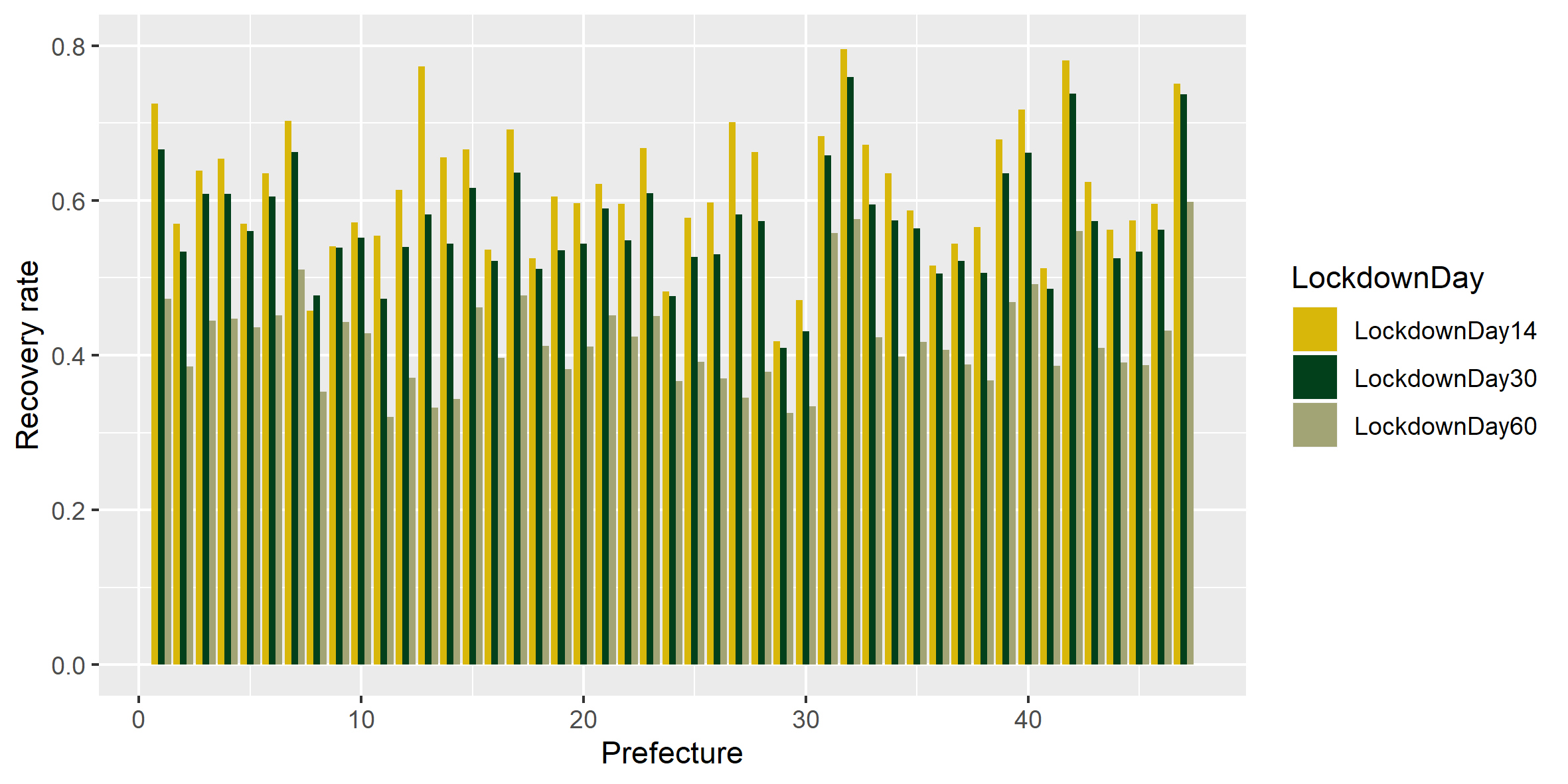}
\caption{Recovery rate in GRP by lifting the lockdown in each prefecture. The recovery rate is defined as the ratio of the increase in the GRP of each prefecture by lifting its lockdown to the decrease in its GRP by all prefectures' lockdowns. The horizontal axis indicates the JIS codes of the prefectures. The yellow, dark green, and light green bars show the recovery rate when lockdowns persist for 14, 30, and 60 days, respectively. 
}
\label{fig:onoffeach}
\end{figure}

Figure~\ref{fig:onoffeach} illustrates the ratio of the increase in the value added production, or gross regional product (GRP), of each prefecture by lifting its lockdown to the decrease in its GRP by all prefectures' lockdowns, which is shown in Figure~\ref{fig:hmap}.

\subsection{Regression analyses}
\label{ch:reg}

In Section \ref{ch:lift}, we conducted regression analyses to examine what attributes of prefectures cause a larger economic recovery by lifting the lockdown in only one prefecture, using Ordinary Least Squares (OLS) models. Table~\ref{tbl:OneLift_corr} shows the correlation coefficients between all the variables used in the regression analysis and Table~\ref{tbl:OneLift_reg} presents the detailed regression results. 

{\footnotesize
\begin{table}[htbp]\centering \caption{Correlation matrix of the variables used in Section \ref{ch:lift}. The definitions of the variables are as follows. 
RecRatio: the recovery rate defined as the ratio of the increase in the GRP of each prefecture by lifting its lockdown to the decrease in its GRP by all prefectures' lockdowns. 
GRP: gross regional product (log). 
Links: the degree (log). 
InLink: the share of links within the prefecture to its all links. 
InLoop: the share of loop flows within the prefecture to its all flows. 
OutLink: the share of outward inter-prefectural links to all the links of the prefecture. 
Potential: the average HH potential of the firms in the prefecture. 
Sub: the share of substitutable suppliers to all suppliers of the prefecture located outside the prefecture. 
\label{tbl:OneLift_corr}}
\begin{tabular}{l  c  c  c  c  c  c  c  c }\hline\hline
\multicolumn{1}{c}{Variable} &RecRatio&GRP&Degree&InLink&InLoop&OutLink&Potential&Sub\\ \hline
RecRatio&1.000\\
GRP&0.311&1.000\\
Degree&0.370&0.965&1.000\\
InLink&0.218&-0.467&-0.374&1.000\\
InLoop&0.432&0.072&0.151&0.720&1.000\\
OutLink&-0.046&0.676&0.661&-0.688&-0.351&1.000\\
Potential&-0.321&0.104&0.090&-0.046&-0.076&0.193&1.000\\
Sub&0.449&0.803&0.829&-0.246&0.307&0.573&0.096&1.000\\
\hline \hline 
\end{tabular}
\end{table}
}

\begin{landscape}
{\small
\begin{table}[htbp]\centering
\caption{Regression results for Section \ref{ch:lift}. The dependent variable is the recovery rate. See the caption of Table~\ref{tbl:OneLift_corr} for the definitions of the independent variables. Standard errors are in parentheses. *** p$<$0.01, ** p$<$0.05, * p$<$0.1.
\label{tbl:OneLift_reg}}
\begin{tabular}{lccccccc} \hline
 & (1) & (2) & (3) & (4) & (5) & (6) & (7) \\
 \hline
 &  &  &  &  &  &  &  \\
InLink &  & 0.426*** &  &  &  &  & 0.401 \\
 &  & (0.133) &  &  &  &  & (0.242) \\
InLoop &  &  & 0.686*** &  &  &  & -0.345 \\
 &  &  & (0.216) &  &  &  & (0.425) \\
OutLink &  &  &  & -0.639** &  &  & -0.375 \\
 &  &  &  & (0.245) &  &  & (0.278) \\
Potential &  &  &  &  & -0.540** &  & -0.527*** \\
 &  &  &  &  & (0.202) &  & (0.179) \\
Sub &  &  &  &  &  & 0.684** & 0.709** \\
 &  &  &  &  &  & (0.275) & (0.283) \\
GRP & 0.0261** & 0.0443*** & 0.0236** & 0.0528*** & 0.0292** & -0.0115 & 0.0242 \\
 & (0.0119) & (0.0122) & (0.0109) & (0.0152) & (0.0112) & (0.0189) & (0.0191) \\
Constant & 0.572*** & 0.285*** & 0.424*** & 0.816*** & 0.565*** & 0.507*** & 0.445** \\
 & (0.0225) & (0.0924) & (0.0512) & (0.0957) & (0.0213) & (0.0338) & (0.182) \\
 &  &  &  &  &  &  &  \\ \hline
Observations & 47 & 47 & 47 & 47 & 47 & 47 & 47 \\
 R-squared & 0.097 & 0.267 & 0.265 & 0.218 & 0.223 & 0.208 & 0.481 \\ \hline
\end{tabular}

\end{table}
}
\end{landscape}

In Section \ref{ch:lifttwo}, we conducted regression analyses to examine what attributes of prefectures cause a larger economic recovery by lifting the lockdown in two prefectures simultaneously, using OLS models. The relative recovery measure defined as the ratio of the increase in the GRP of prefecture $a$ when it lifts its lockdown together with prefecture $b$ to its increase when prefecture $a$ lifts its lockdown alone. Table~\ref{tbl:DoubleLift_corr} shows the correlation coefficients between all the variables used in the regression analysis and Table~\ref{tbl:DoubleLift_reg} presents the detailed regression results. 

\begin{figure}[htb]
\centering
\includegraphics[width=.7\linewidth]{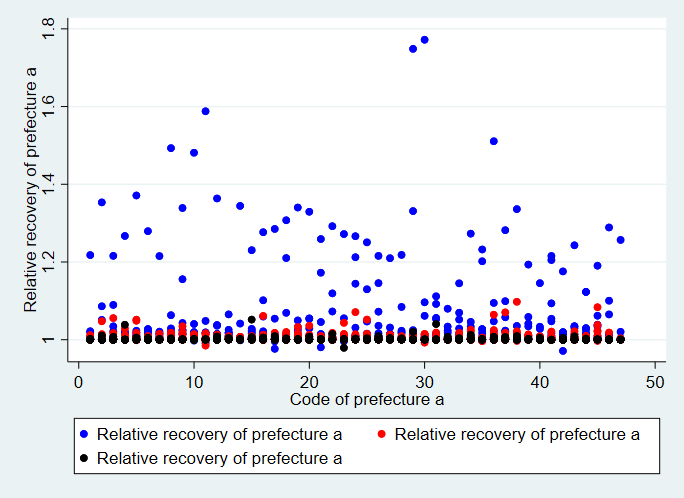}
\caption{Relative recovery from lifting the lockdown together to the recovery from lifting the lockdown alone. The relative recovery measure is defined as the ratio of the increase in the GRP of prefecture $a$ when it lifts its lockdown together with prefecture $b$ to its increase when prefecture $a$ lifts its lockdown alone. The horizontal axis shows the JIS code of prefecture $a$. The colour of each dot indicates whether the GRP of prefecture $b$ is among the top 10 (blue), the bottom 10 (black), or others (red). }
\label{fig:DoubleLift_prefcode}
\end{figure}

{\footnotesize
\begin{table}[htbp]\centering \caption{Correlation matrix of the variables used in Section \ref{ch:lifttwo}. The definitions of the variables are as follows. 
$Recov_a$: the relative recovery of prefecture $a$ defined as the ratio of the increase in the GRP of prefecture $a$ by lifting its lockdown together with prefecture $b$ to its increase by lifting its lockdown alone. 
$Link_{ab}$: the share of links from $a$ to $b$ to all links from $a$.
$Link_{ba}$: the share of links from $b$ to $a$ to all links from $a$. 
$Pot_{ab}$: the share of potential flows from $b$ to $a$ to the total links of $a$.
$Pot_{ba}$: the share of potential flows from $a$ to $b$ to the total links of $a$.
$Sub_{ab}$: the share of suppliers substitutable by those in $b$ to $a$'s suppliers outside $a$ and $b$. 
$Sub_{ba}$: the share of suppliers substitutable by those in $a$ to $b$'s suppliers outside $a$ and $b$. 
$Loop_{ab}$: the share of loop flows between $a$ and $b$ to the total flows between the two.
$Bi_{ab}$: the number of inter-prefecture links between $a$ and $b$ in logs. 
$GRP_{j}$: GRP of $b$ in logs.\label{tbl:DoubleLift_corr}}
\begin{tabular}{l  c  c  c  c  c  c  c  c  c  c }\hline\hline
\multicolumn{1}{c}{Variable} &$Recov_a$&$Link_{ab}$&$ Link_{ba}$&$ Pot_{ab}$&$ Pot_{ba}$&$ Sub_{ab}$&$ Sub_{ba}$&$ Loop_{ab}$&$ Bi_{ab}$&$ GRP_b$\\ \hline
$ Recov_a$&1.000\\
$ Link_{ab}$&0.820&1.000\\
$ Link_{ba}$&0.818&0.966&1.000\\
$ Pot_{ab}$&0.870&0.927&0.961&1.000\\
$ Pot_{ba}$&0.808&0.915&0.955&0.968&1.000\\
$ Sub_{ab}$&0.071&0.185&0.238&0.182&0.243&1.000\\
$ Sub_{ba}$&0.813&0.961&0.966&0.946&0.948&0.237&1.000\\
$ Loop_{ab}$&0.879&0.911&0.952&0.986&0.979&0.206&0.940&1.000\\
$ Bi_{ab}$&0.392&0.543&0.564&0.499&0.528&0.572&0.572&0.504&1.000\\
$ GRP_b$&0.563&0.610&0.597&0.602&0.582&0.056&0.643&0.596&0.576&1.000\\
\hline \hline 
 \end{tabular}
\end{table}
}

\begin{landscape}
{\small 
\begin{table}[htbp]
\centering
\caption{Regression results for Section \ref{ch:lifttwo}. The dependent variable is the relative recovery measure. See the caption of Table~\ref{tbl:DoubleLift_corr} for the definitions of the independent variables. Standard errors are in parentheses. *** p$<$0.01, ** p$<$0.05, * p$<$0.1.}
\label{tbl:DoubleLift_reg}
\begin{tabular}{lccccccc} \hline
 & (1) & (2) & (3) & (4) & (5) & (6) & (7) \\
\hline
 &  &  &  &  &  &  &  \\
$Link_{ab}$ & 0.519*** &  &  &  &  &  & 0.440*** \\
 & (0.0175) &  &  &  &  &  & (0.0306) \\
$Link_{ba}$ &  & 0.619*** &  &  &  &  & -0.375*** \\
 &  & (0.0199) &  &  &  &  & (0.0460) \\
$Pot_{ab}$ &  &  & 8.333*** &  &  &  & 0.277 \\
 &  &  & (0.198) &  &  &  & (0.624) \\
$Pot_{ba}$ &  &  &  & 8.076*** &  &  & -17.82*** \\
 &  &  &  & (0.271) &  &  & (0.644) \\
$Loop_{ab}$ &  &  &  &  & 3.841*** &  & 10.06*** \\
 &  &  &  &  & (0.0844) &  & (0.309) \\
$Sub_{ba}$ &  &  &  &  &  & 1.564*** & -0.248** \\
 &  &  &  &  &  & (0.0550) & (0.0989) \\
$Bi_{ab}$ & -0.00211*** & -0.00288*** & -0.00182*** & -0.00174*** & -0.00225*** & -0.00235*** & -0.000602** \\
 & (0.000413) & (0.000415) & (0.000352) & (0.000407) & (0.000341) & (0.000423) & (0.000298) \\
$GRP_b$ & -0.0186*** & -0.0186*** & -0.0135*** & -0.0232*** & -0.0120*** & -0.0220*** & -0.00540*** \\
 & (0.00210) & (0.00205) & (0.00181) & (0.00202) & (0.00175) & (0.00208) & (0.00146) \\
$GRP_b^2$ & 0.00652*** & 0.00676*** & 0.00467*** & 0.00795*** & 0.00431*** & 0.00712*** & 0.00192*** \\
 & (0.000490) & (0.000471) & (0.000422) & (0.000458) & (0.000405) & (0.000489) & (0.000350) \\
Constant & 1.019*** & 1.023*** & 1.016*** & 1.021*** & 1.017*** & 1.024*** & 1.006*** \\
 & (0.00236) & (0.00233) & (0.00208) & (0.00236) & (0.00200) & (0.00239) & (0.00168) \\
 &  &  &  &  &  &  &  \\ \hline
Observations & 2,162 & 2,162 & 2,162 & 2,162 & 2,162 & 2,162 & 2,162 \\
 R-squared & 0.713 & 0.721 & 0.778 & 0.714 & 0.794 & 0.706 & 0.865 \\ \hline
\end{tabular}

\end{table}

}
\end{landscape}

\begin{figure}[htb]
\centering
\includegraphics[width=.75\linewidth]{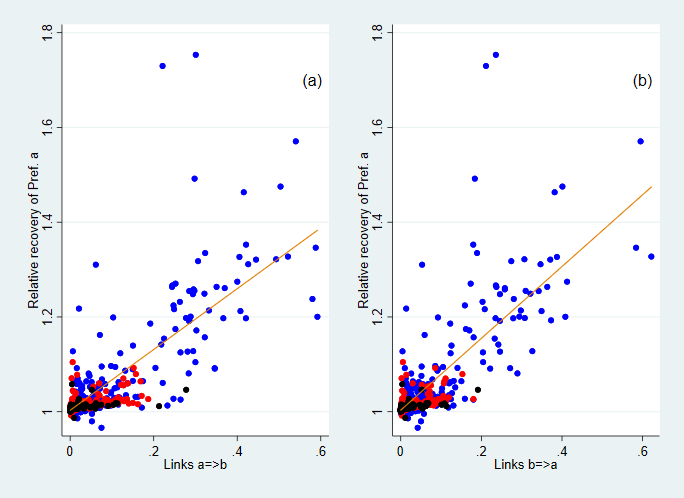}
\caption{Correlation between the relative recovery and selected network measures. The vertical axis indicates the relative recovery of prefecture $a$, defined as the ratio of the increase in the GRP of prefecture $a$ by lifting its lockdown together with prefecture $b$ to its increase by lifting its lockdown alone. The effect of the GRP of $b$ and total links between the two are excluded from the relative recovery measure. The variable in the horizontal axis is given by 
Equations \ref{eq:link_ab} and \ref{eq:link_ba} in panels (a) and (b), respectively. 
The orange line in each panel signifies the fitted value from a linear regression that controls for the effect of the GRP of $b$ and total number of links between $a$ and $b$.
The blue, black, and red dots show the pairs of prefectures $a$ and $b$ for which the GRP of $b$ is among the top 10, bottom 10, and others, respectively.}
\label{fig:DoubleLift_comb2}
\end{figure}

\begin{figure}[htb]
\centering
\includegraphics[width=\linewidth]{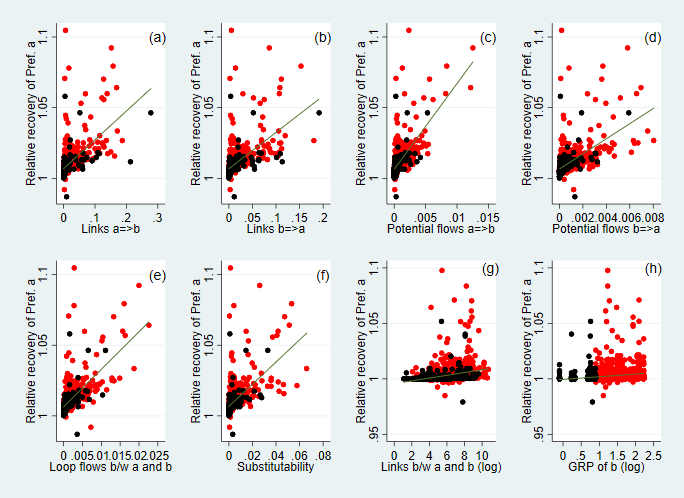}
\caption{Correlation between the relative recovery and selected network measures. See the caption of Figures \ref{fig:DoubleLift_comb} and \ref{fig:DoubleLift_comb2} for the definitions of the variables used here. The green line in each panel signifies the fitted value from a linear regression that controls for the effect of the GRP of $b$ and total number of links between $a$ and $b$ in (a)--(g).
The black and red dots show the pairs of prefectures $a$ and $b$ for which the GRP of $b$ is among the bottom 10 and between 11 and 37, respectively.}
\label{fig:DoubleLift_comb_notop10}
\end{figure}

\begin{figure}[htb]
\centering
\includegraphics[width=\linewidth]{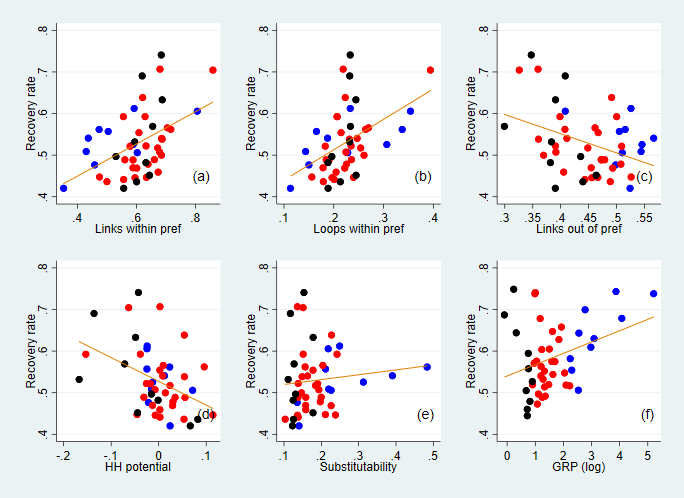}
\caption{Correlation between the recovery rate and selected network measures. See the caption of Figure~\ref{fig:OneLift_comb} for the definitions of the variables used here. The orange line in each panel specifies the fitted value from a linear regression that controls for the effect of GRP in (b)--(f).
The blue, black, and red dots show the prefectures whose GRP is among the top 10, the bottom 10, or others, respectively.}
\label{fig:OneLift_comb_heavy}
\end{figure}

\begin{figure}[htb]
\centering
\includegraphics[width=\linewidth]{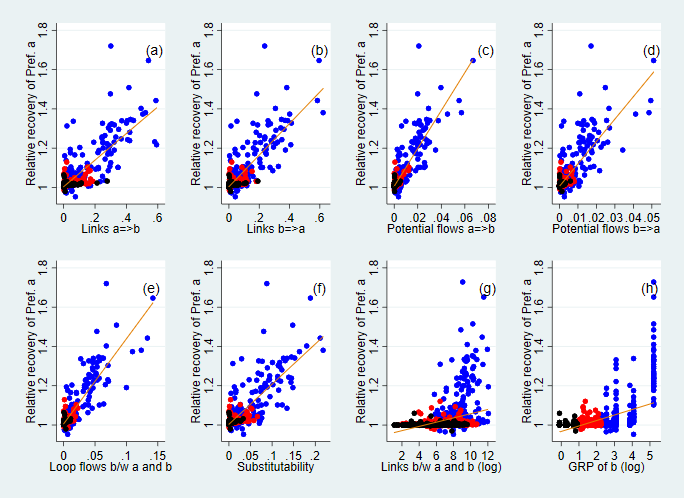}
\caption{Correlation between the relative recovery and selected network measures. See the caption of Figures \ref{fig:DoubleLift_comb} \ref{fig:DoubleLift_comb} for the definitions of the variables used here. The red line in each panel signifies the fitted value from a linear regression that controls for the effect of the GRP of $b$ and total number of links between $a$ and $b$ in (a)--(g).
The blue, black, and red dots show the pairs of prefectures $a$ and $b$ for which the GRP of $b$ is among the top 10, the bottom 10, or others, respectively.}
\label{fig:DoubleLift_comb_heavy}
\end{figure}

To check the robustness of our main results, we experimented with different rates of reduction in production capacity, where we assume the share of working from home is zero for all the sectors in Supplementary Information~Table~\ref{ch:strength}. In other words, in this alternative simulation analysis, we assume a stricter level of lockdown. Supplementary Information~Figures \ref{fig:OneLift_comb_heavy} and \ref{fig:DoubleLift_comb_heavy} present the results, which are essentially the same as our benchmark results in Figures \ref{fig:OneLift_comb} and \ref{fig:DoubleLift_comb}.

\end{document}